\documentclass{elsarticle}
\usepackage{mathtools}
\usepackage{times}
\usepackage{latexsym}
\usepackage{multicol}
\usepackage{graphicx}
\usepackage{algorithm}
\usepackage[noend]{algpseudocode}
\usepackage{varwidth}
\usepackage{multirow}
\usepackage{hyperref}

\title{Topology for Substrate Routing \\ in Semiconductor Package Design}

\author[1]{Rak-Kyeong Seong \corref{cor1}\fnref{fn1}}

\author[1]{Jaeho Yang}
\author[1]{Sang-Hoon Han}

\cortext[cor1]{Corresponding author: rk.seong@samsung.com}
\address[1]{Samsung SDS,  AI Advanced Research Lab,
  Samsung R\&D Campus, Seocho-Gu, Seoul, South Korea}

\begin{document}

\begin{abstract}
In this work, we propose a new signal routing method for solving routing problems that occur in the design process of semiconductor package substrates. 
Our work uses a topological transformation of the layers of the package substrate in order to simplify the routing problem into a problem of connecting points on a circle with non-intersecting straight line segments. 
The circle, which we call the Circular Frame, is a polygonal schema, which is originally used in topology to study the topological structure of 2-manifolds. 
We show through experiments that our new routing method based on the Circular Frame competes with certain grid-based routing algorithms.
\end{abstract}  
\maketitle

\section{Introduction}

\begin{figure}[h!]
  \centering
  \includegraphics[trim={0cm 0cm 0cm 0cm}, width=0.7\linewidth]{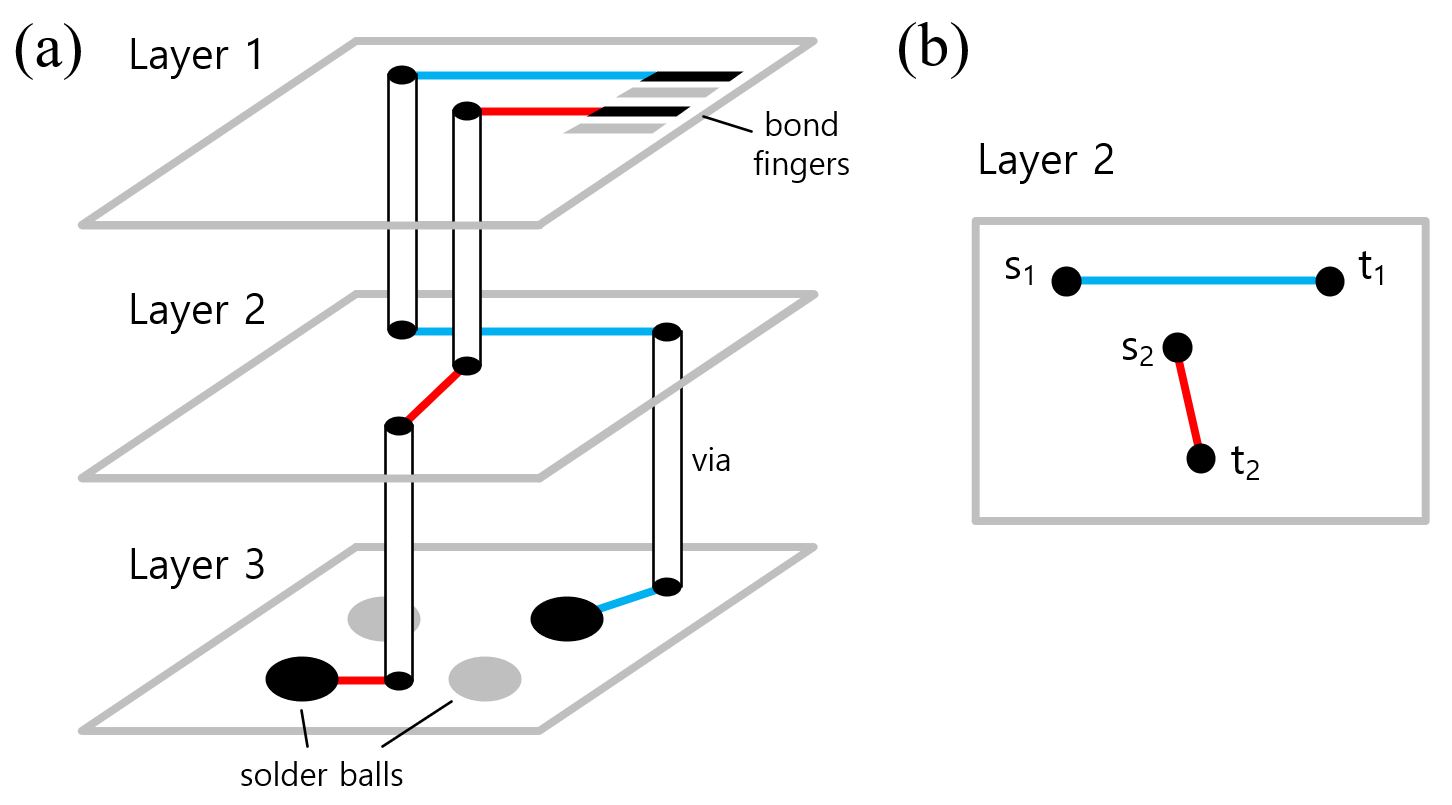}
  \caption{
{\bf Fine Pitch Ball Grid Array (FBGA) Package Substrate Layout.} (a) An illustration of a 3-layered FBGA package substrate with vias connecting different substrate layers. (b) Each individual layer (here layer 2) has its own set of start and end points that need to be connected with non-intersecting paths. 
  \label{fig:f01}}
\end{figure}

Semiconductor devices are at the forefront of innovation in the information technology (IT) industry and play an essential role in driving innovations in areas such as consumer electronics, telecommunications, artificial intelligence or data analysis and security.  Although semiconductor devices play such a pivotal role in IT innovation, the integrated circuit (IC) packaging process of semiconductor devices still heavily relies on human expertise. For substrates in, for example, chip-scale packages such as multi-layered Fine Pitched Ball Grid Array (FBGA) packages as illustrated in Fig. \ref{fig:f01}, most of the design process is about finding the optimal connections between bond fingers, vias and solder balls. Given the variety of types for semiconductor packages, the problem of substrate routing is challenging. As a result, substrate routing problems are often solved with the help of routing methods that are implemented in many computer-aided design (CAD) solutions. In line with recent advances in Electronic Design Automation (EDA), in this work, we outline a new routing method for package substrate design that competes with the performance of other routing methods.

\begin{figure}[h!]
  \centering
  \includegraphics[trim={0cm 0cm 0cm 0cm}, width=0.7\linewidth]{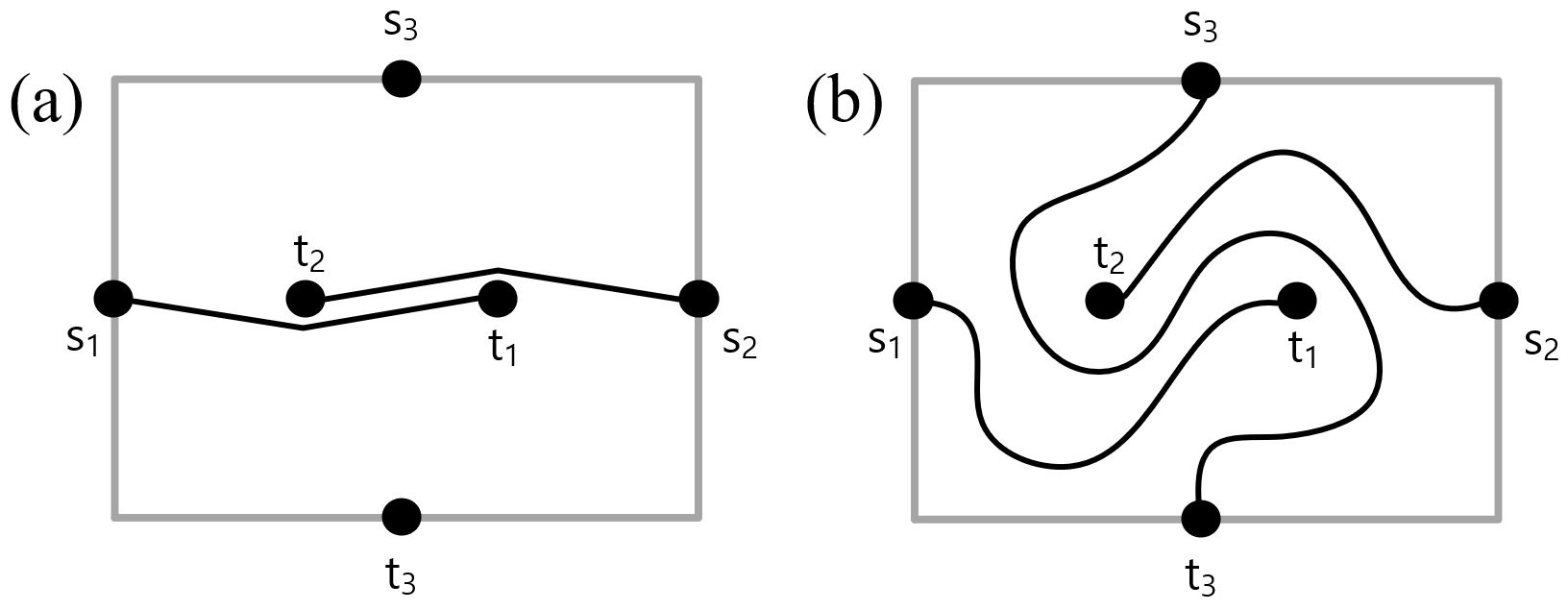}
  \caption{
{\bf Geometrical and Topological Routers.} (a) In geometrical routers, start ($s_i$) and end ($t_i$) points are sequentially connected with shortest paths, which can result in a lack of clearance for any following pairs, in this case $s_3$ and $t_3$. (b) In topological routers, the connection problem only deals with relative positions, avoiding problems of clearance.
  \label{fig:f02}}
\end{figure}

The problem of finding non-intersecting paths that connect a set of start and end points on a plane is one of the oldest problems in computational geometry and graph theory. We know that Dijkstra's algorithm and the A*-algorithm \cite{dijkstra1959note, hart1968formal} are examples of graph traversal algorithms, which are used to solve such routing problems. However, substrate routing becomes exponentially more complicated with an increasing number of start and end point pairs.

Most routing algorithms such as Dijkstra's algorithm, the A*-algorithm and other grid-based Maze Router algorithms \cite{5219222, 1600289, 285679, 920691, 10.1145/267665.267686} are known as \textit{geometrical routers}. Their disadvantage is that when start and end point pairs are connected sequentially on consecutive shortest paths, it becomes increasingly more likely that there will be not enough clearance left for consecutive connections between pairs of points. This problem with geometrical routers is illustrated in Fig. \ref{fig:f02}.   

In this work, we are interested in a different class of routers known as \textit{topological routers} \cite{Dai}. In order to connect fully all points, topological routers aim to find the \textit{topological class} of the connections first, \textit{i.e.} the relative positions of paths. After the topological class of the connecting paths is found, with a choice of representation scheme, absolute coordinates are assigned to represent the routing result in real space. This avoids situations where there is a lack of clearance as it is often the case for geometrical routers. For topological routers, paths can always be inserted between already routed paths in order to solve the connection problem. Fig. \ref{fig:f02} illustrates this difference between geometrical and topological routers. 

Our work is based on the concept of topological routers and proposes a novel topological representation and routing algorithm for substrate routing that competes with the performance of conventional geometrical routers. We make use of topology, more specifically the study of 2-manifolds and polygonal schema \cite{fulton, papadopoulou1996k,efrat2006computing, erickson2011shortest} in mathematics in order to topologically transform the package substrate into a simpler abstract environment where routing design can be performed more straightforwardly. 

In an earlier work \cite{2021arXiv210503386S}, we outlined the general principle of our new method for general routing problems. In the current work, we extend our proposal with a focus on the problem of substrate routing in semiconductor chip package design. In particular, we apply our substrate routing method to an explicit example of a Fine Pitch Ball Grid Array (FBGA) package. 

Note that our work concentrates on a substrate routing method that finds a fully connected routing solution and does not take into account other metrics such as the wire length or optimal placement of via points.\footnote{Our routing method can be adjusted to take into account an optimization metric and this will be the subject of upcoming work.} Our work also concentrates on signal routing in substrates where the routing problem involves connections between a single start and a single corresponding end point.\footnote{Multi-pin routing that occurs in power and ground routing or plating lines can be covered in a generalized version of our method, which we plan to cover in future works.}

We test our routing method's performance against geometrical routers and conclude with a summary of results and an actual FBGA package substrate design that was completed using our new routing method.

\section{Background}

\begin{figure}[h!]
  \centering
  \includegraphics[trim={0cm 0cm 0cm 0cm}, width=0.65\linewidth]{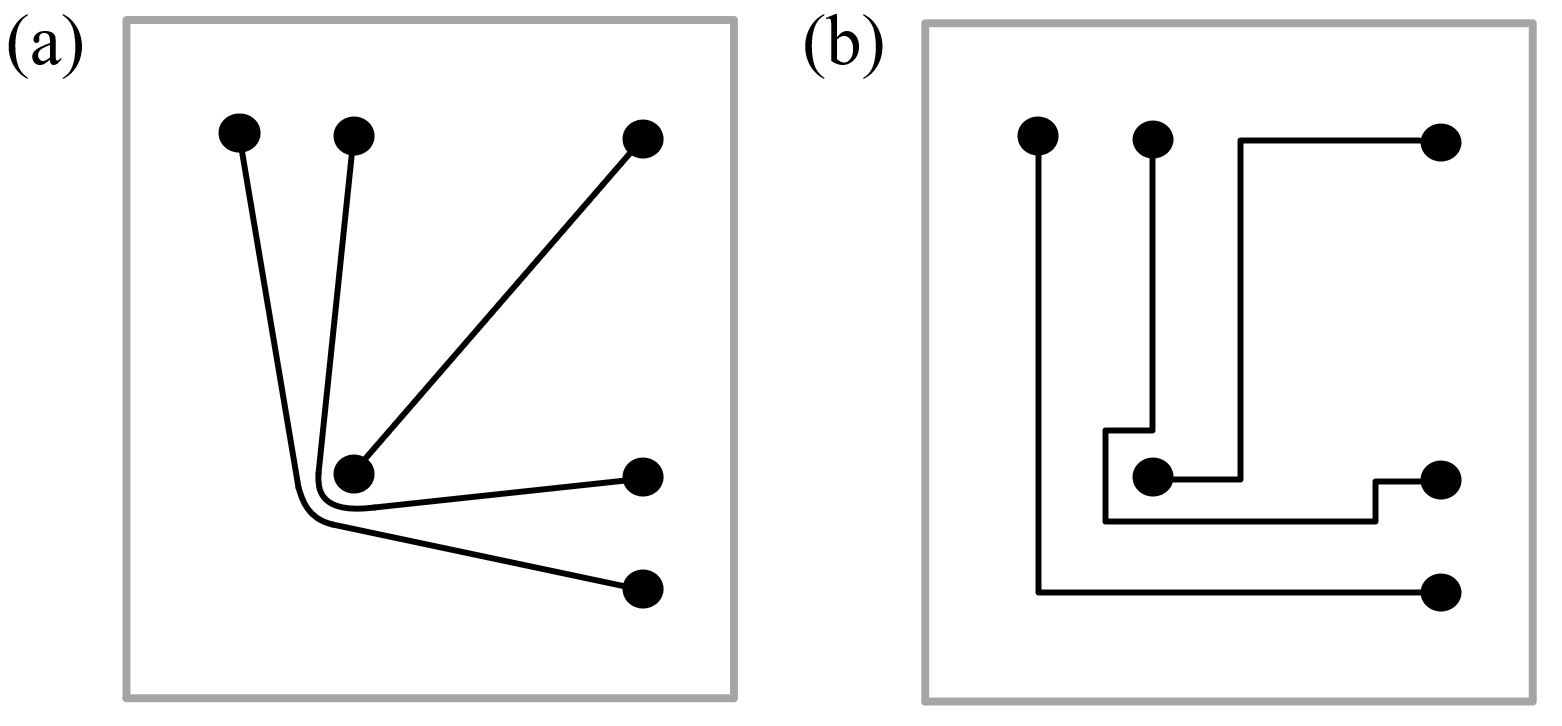}
  \caption{
{\bf Preserving Routing Topology.} (a) Rubber-band sketch representation of a connected set of start and end points, (b) compared to a rectilinear representation of the same connected solution with the same routing topology.  
  \label{fig:f03}}
\end{figure}

In this work, we propose based on our earlier work in \cite{2021arXiv210503386S} a new method of solving routing problems that occur during the package substrate design process by using topology. The idea of making use of concepts in topology for designing circuits is not new as shown by the works on \textit{rubber-band routing} in \cite{Dai, Dai_1991}. These works discuss how certain design features in circuit design can be altered without changing the connections between points, \textit{i.e.} the topology of the paths, as illustrated in Fig. \ref{fig:f03}. Moreover, they give an insight into how paths can be bent and moved in such a way that problems of clearance occurring with traditional geometrical routers can be avoided.

Several routing algorithms have been proposed for EDA since the 1990s \cite{1600289,285679, Dai_1991, 10.1145/267665.267686}, which are based on the idea of grid-dependent geometrical routers. Moreover, more recent work in EDA considers applications and improvements on geometrical routers in areas such as length matching routing \cite{8942123}, escape routing \cite{8648510,8102207, 7827599, Weng2020URBERUR}, routing with obstacle avoidance \cite{8664730} and pin assignment and placement algorithms \cite{8715126}. In comparison, topological routers have been studied less extensively \cite{Dai, Dai_1991} although, as mentioned above, they have considerable advantages over geometrical routers.

\begin{figure}[h!]
  \centering
  \includegraphics[trim={0cm 0cm 0cm 0cm}, width=0.7\linewidth]{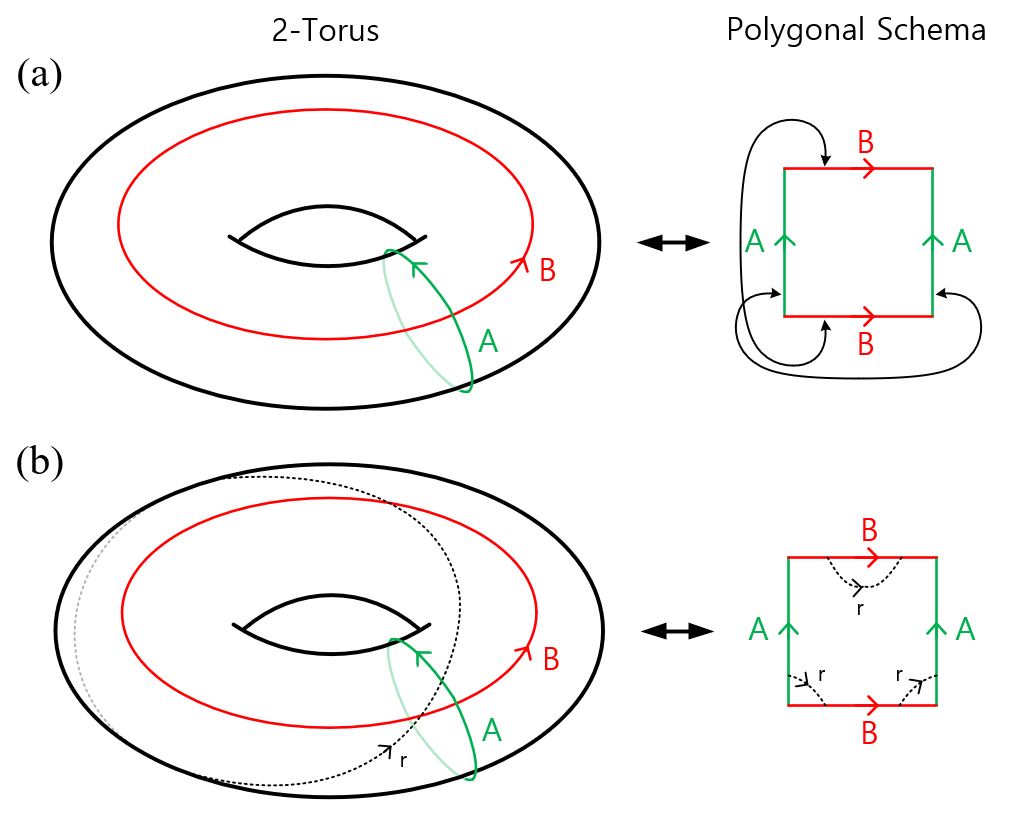}
  \caption{
{\bf Polygonal Schema.} (a) A torus with its corresponding polygonal schema, which is a rectangle with opposite edges identified with each other. (b) A path on the torus can be represented as a path on the corresponding polygonal schema.  
  \label{fig:f04}}
\end{figure}

In contrast to the developments made in geometrical routing, our work tries to push forward the development of topological routing. In particular, our work proposes the use of a novel topological transformation to completely transform the substrate routing environment into a topologically equivalent environment. This is a completely new approach for routing in package substrates. Our proposed transformation maps the routing problem to a topologically simpler space where the problem can be solved more straightforwardly. This is the case when in the new environment only relative positions are preserved under the transformation. Given that the transformation is reversible, after all nets are connected, the space with the routing result is transformed back to its original substrate environment.

Such topological transformations and representations that preserve relative positions rather than absolute positions occur extensively in the study of compact 2-manifolds through \textit{polygonal schema} \cite{fulton}. These were introduced in mathematics to study the topology of compact 2-manifolds and are particularly useful in representing the \textit{homotopy} of paths on these manifolds \cite{efrat2006computing}. As a result, polygonal schema appeared also extensively in relation to so-called non-crossing walk problems on compact 2-manifolds \cite{papadopoulou1996k, erickson2011shortest}.

Let us illustrate briefly the concept behind polygonal schema using one of the simplest compact 2-manifolds, the Riemann surface of genus 1, which is also known as a torus or doughnut. The torus can be represented by a rectangle when opposite sides of the rectangle are identified with each other. Any such simple convex polygon together with a boundary gluing pattern shown in Fig. \ref{fig:f04} is known as a polygonal schema of the represented 2-manifold. Using the example of the 2-torus, we learn that a rectangle with its opposite boundary sides identified with each other is topologically equivalent to a torus. We can see from this example that even though a torus is 3-dimensional, it can be much more straightforwardly represented by its 2-dimensional polygonal schema. 

\begin{figure}[h!]
  \centering
  \includegraphics[trim={0cm 0cm 0cm 0cm}, width=1\linewidth]{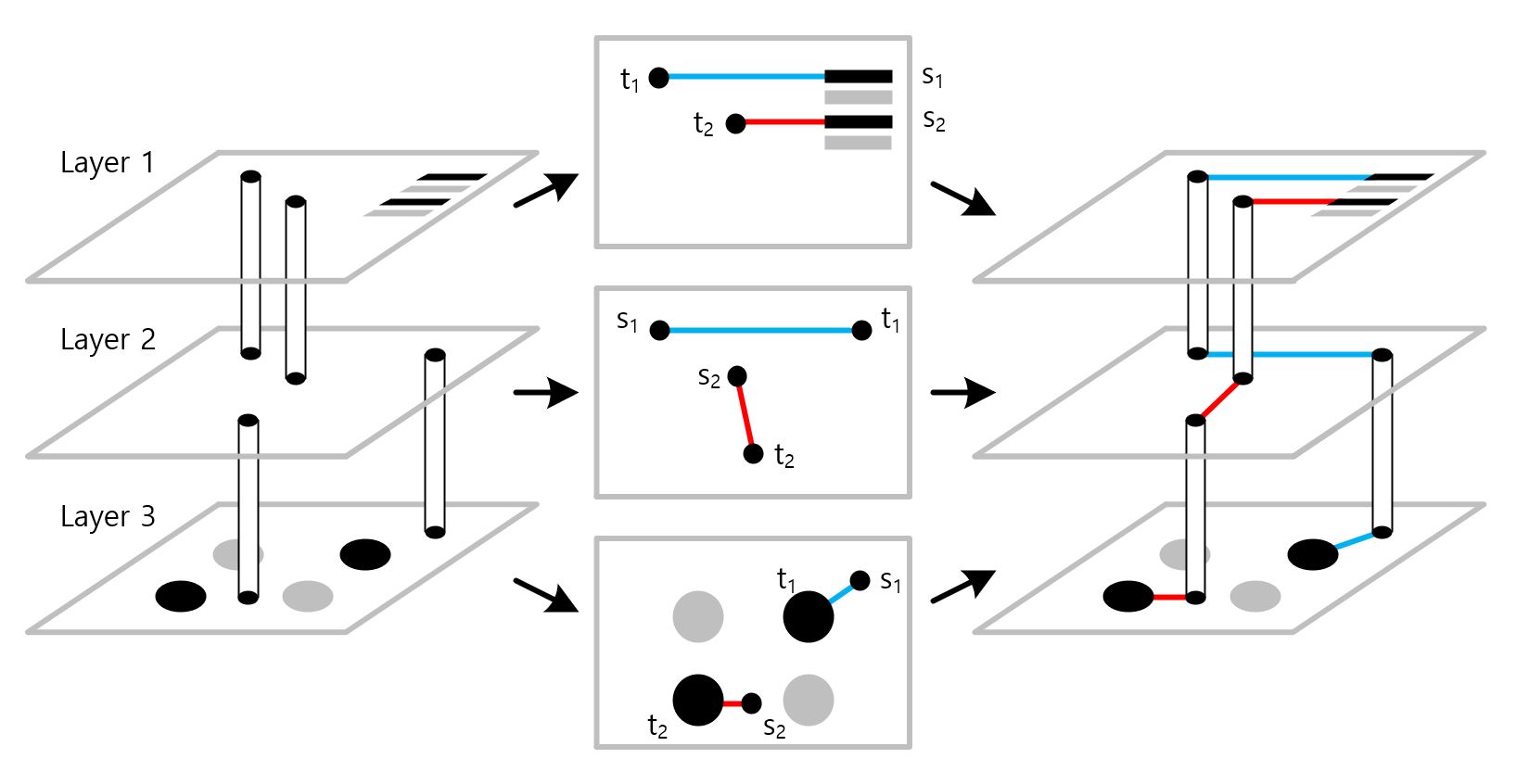}
  \caption{
{\bf Routing Problem in a multi-layered FBGA Package Substrate.} Each layer of the package substrate has its own set of start and end points. After solving the routing problem on each substrate layer, the layers can be connected again along the vias.  
  \label{fig:f05}}
\end{figure}

We claim that a semiconductor package substrate, which usually contains multiple interconnected layers, can be described topologically in terms of polygonal schema. Substrate layers, which are connected by vias, can be separated and individually represented by polygonal schema. Because we split the layers for the topological transformation, each layer has its layer-specific start and end points corresponding to either pins, solder balls or vias. We keep track of which via connects which layers together so that when we reverse the topological transformation, we are able to sew back together the vias between each pair of layers to form the original multi-layered package substrate as shown in Fig. \ref{fig:f05}. Note that the locations of the via points plays an important role in the overall global routing solution. Since we focus on the problem of finding a fully connected routing solution and consider no other optimization metrics, we refer to future work on optimizing the routing solution using our method.

In the following section, we describe how we make use of the topological transformation specific to our problem and describe a method of how to complete the routing in the topologically transformed routing environment.

\begin{figure}[h!]
  \centering
  \includegraphics[trim={0cm 0cm 0cm 0cm}, width=0.65\linewidth]{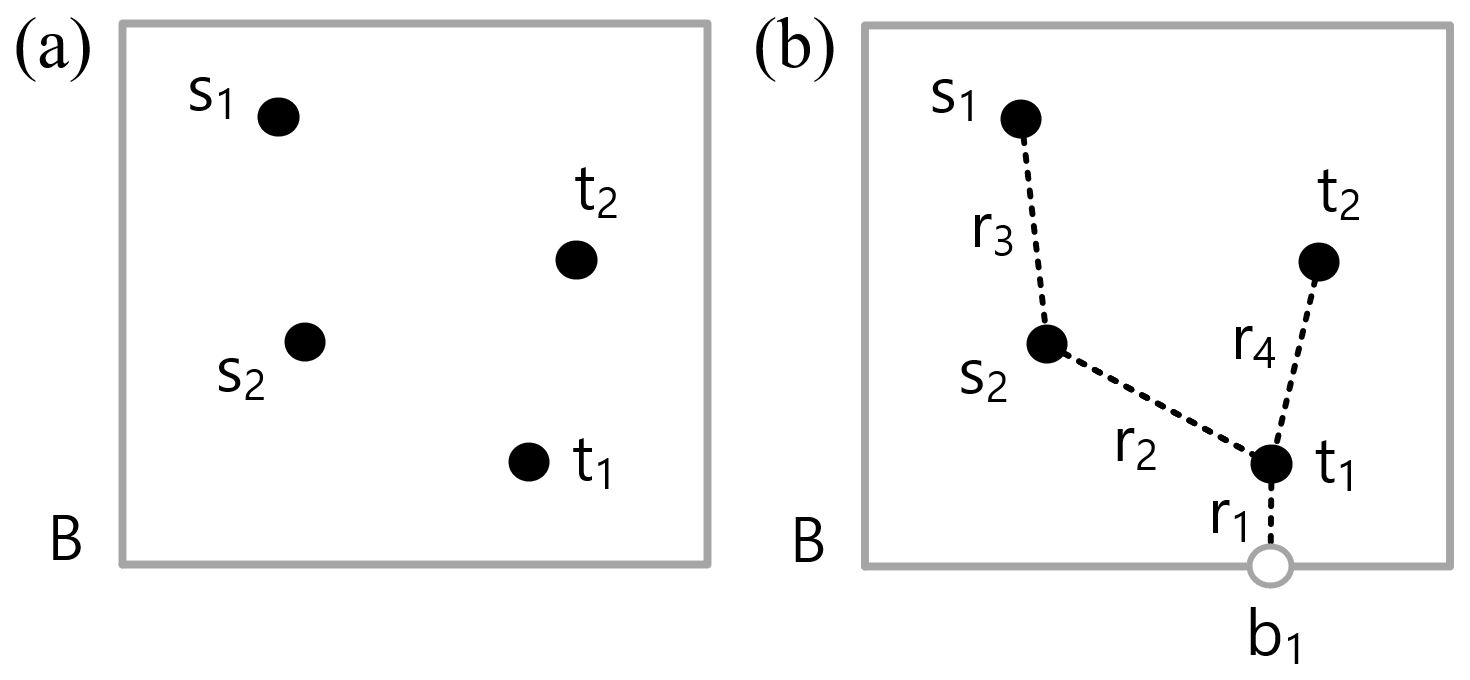}
  \caption{
{\bf Start Points, End Points and Trees.} (a) Start points $s_i\in S$ and end points $t_i \in T$ on a plane bounded by $B$. (b) Trees $R$ made of edges $r_i$ connect all $s_i$ and $t_i$ to  points $b_j$ on the boundary $B$. 
  \label{fig:f06}}
\end{figure}

\section{Circular Frame \label{sec:circularframe}}

Let there be a set $S$ of start points $s_i$ and a set $T$ of end points $t_i$ with pairwise identification $s_i\rightarrow t_i$. For our routing problem, we call such a pair a \textit{net}. These points are on a plane bounded by $B$ as shown in Fig. \ref{fig:f06} (a). In order to transform this environment, we introduce trees $R$ consisting of a set of edges $r_i$ such that these edges have at their ends either $s_i\in S$, $t_i\in T$ or $b_i\in B$. All points in $S$ and $T$ are each connected to a single tree. Note that a tree $R$ is always connected by exactly one edge with the boundary $B$ at a point $b_i$ as shown in Fig. \ref{fig:f06} (b). 
These trees $R$ can be found using a minimum spanning tree algorithm such as Kruskal's algorithm.\footnote{Note that the choice of method for finding the spanning trees may lead to a single tree. Furthermore, the choice will impact the routing result and leads to questions about optimization that will be studied in future work \cite{6461975}.} Such an algorithm needs to be generalized such that each tree $R$ gets connected to the boundary $B$ at a point $b_i$ by a single edge $r_i$. The start and end points do not need to be connected to $B$ by a single tree $R$. Each point can be connected to the boundary $B$ by separate trees where each tree is separately connected to $B$.

\begin{figure}[h!]
  \centering
  \includegraphics[trim={0cm 0.5cm 0cm 0cm}, width=0.7\linewidth]{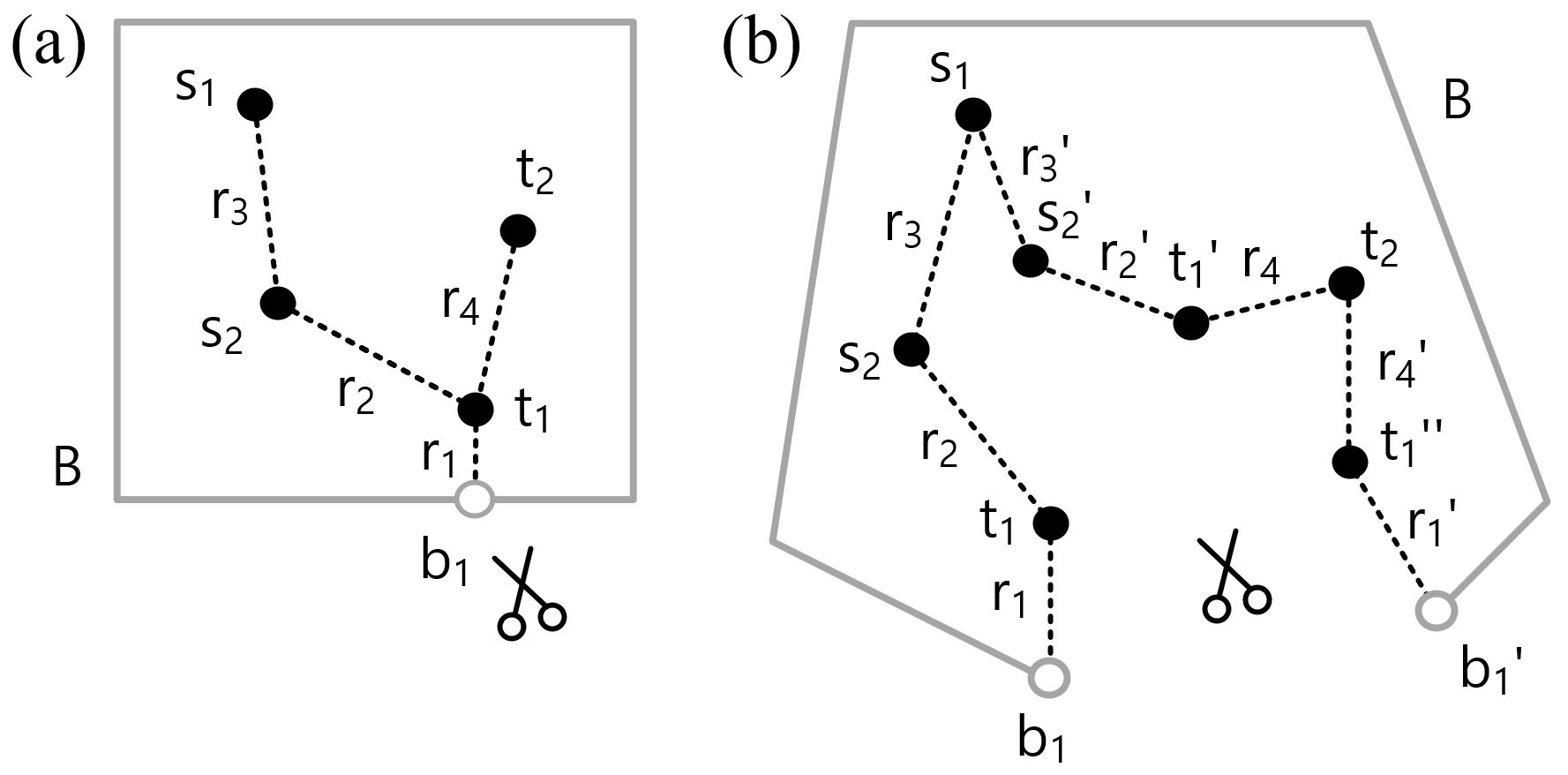}
  \caption{
{\bf Cutting along Trees.} (a) Cutting the plane along the tree edges $r_i\in R$ (b) splits the points connected to the edges. 
  \label{fig:f08}}
\end{figure}

Our proposed topological transformation cuts the plane along all the edges $r_i$ such that all points in $S$ and $T$ are now placed on a new boundary that includes the cut-lines along $r_i$ as shown in Fig. \ref{fig:f08}. The cutting process splits some of the points $s_i$ and $t_i$ to multiple copies if the original points are connected to more than one tree edge $r_i$. The boundary points at which trees are attached to the original boundary $B$ are always separated into a pair $b_i$ and $b_{i}'$. We also notice that during the cutting process the edges $r_i$ separate into pairs $r_i$ and $r_{i}'$.

\begin{figure}[h!]
  \centering
  \includegraphics[trim={0cm 0.5cm 0cm 0cm}, width=0.7\linewidth]{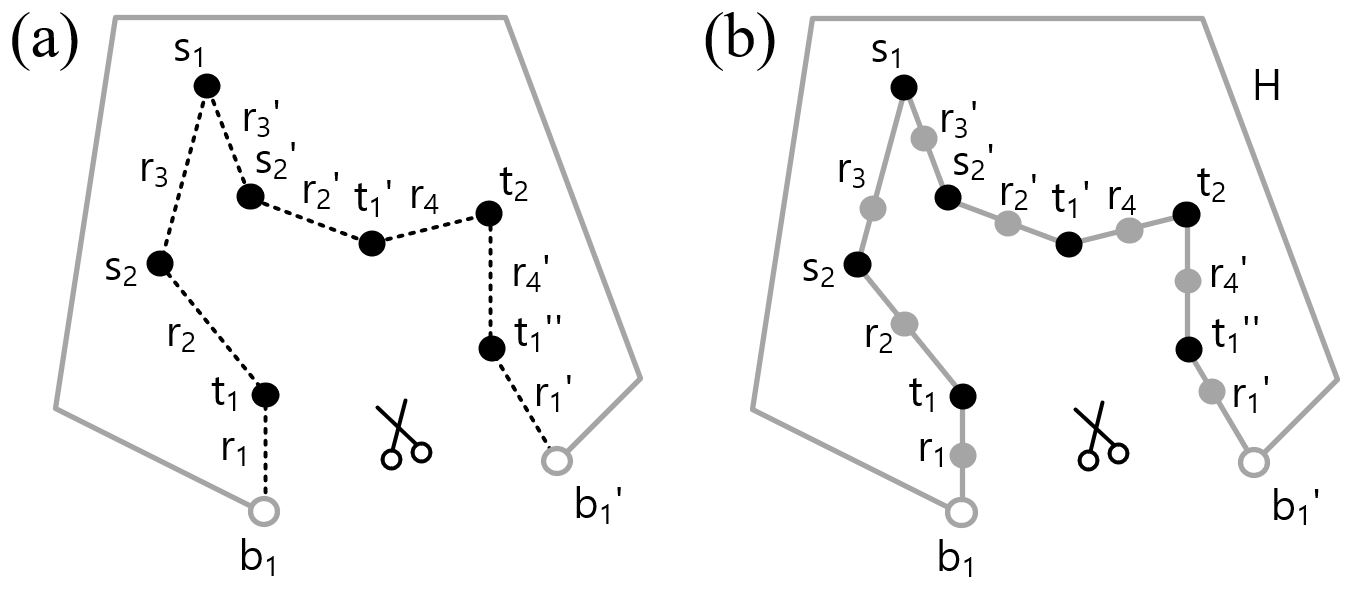}
  \caption{
{\bf Tree Lines as Points.} (a) The cutting process splits the edges into pairs $r_i$, $r_{i}'$. (b) Each of the edges can be represented as points on the combined boundary $H$. All points are now on $H$.  
  \label{fig:f09}}
\end{figure}

We pinch the edges $r_i$ and $r_{i}'$ originating from the trees $R$ in such a way that they are also represented by points on the new boundary $H$ as shown in Fig. \ref{fig:f09}. As a result, the start points $s_i$, end points $t_i$, the tree edges $r_i$, boundary points $b_i$ and their corresponding partners generated by the cutting process are all represented as points on a single combined boundary $H$ as shown in Fig. \ref{fig:f09}.

\begin{figure}[h!]
  \centering
  \includegraphics[trim={0cm 0.5cm 0cm 0cm}, width=0.8\linewidth]{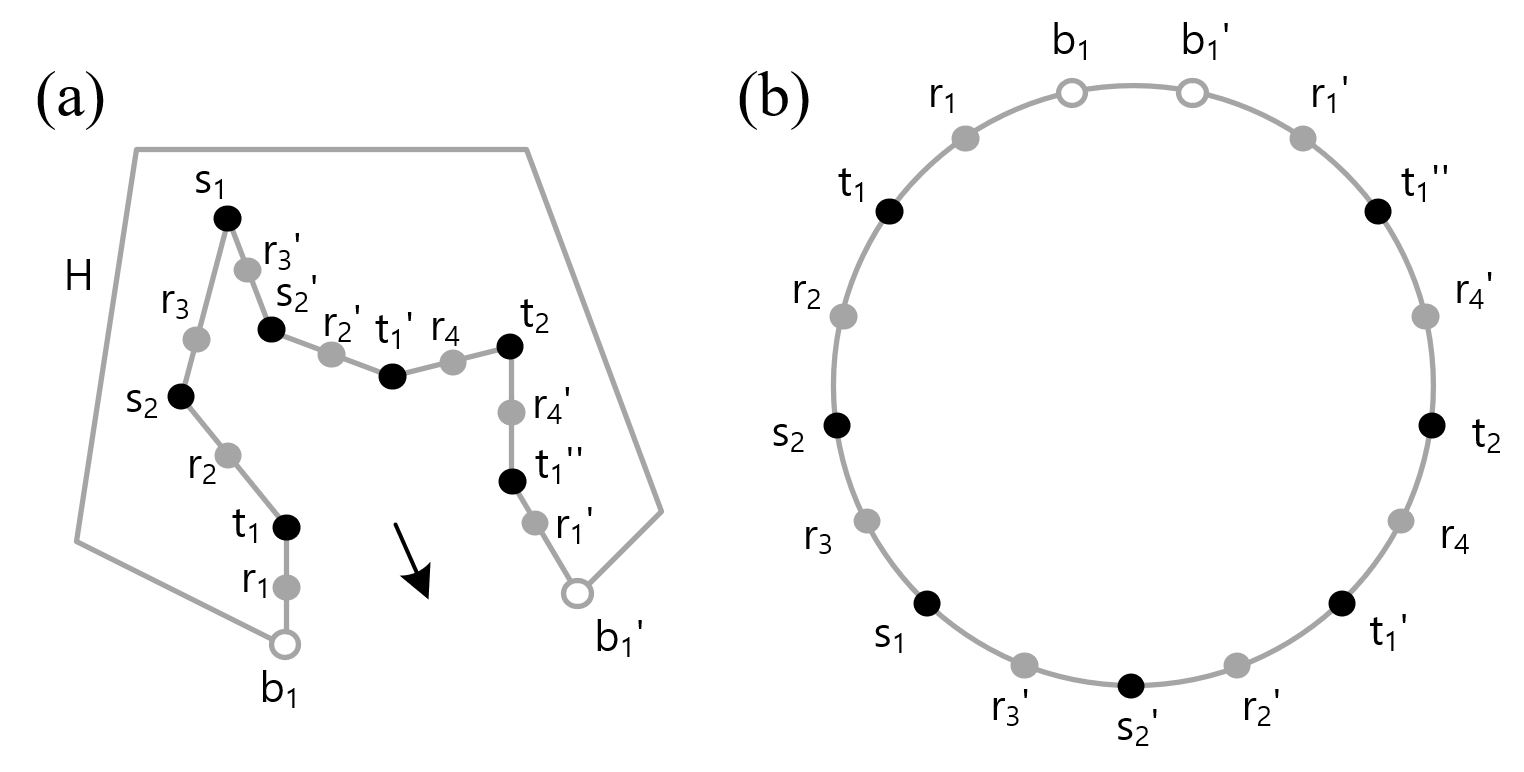}
  \caption{
{\bf Circular Frame.} (a) The combined boundary $H$ can be deformed to form (b) a circle. The interior of the circle represents the original substrate layer that was cut, and the start, end, boundary and tree edge points are all on the circle. We call this representation of the original substrate layer the \textit{Circular Frame}.   
  \label{fig:f10}}
\end{figure}

The combined boundary $H$ can be deformed into a circle as illustrated in Fig. \ref{fig:f10}. We call this representation of the original substrate layer the \textit{Circular Frame}. The order in which the points appear along the circle is the same as they appear when one traverses $H$ in a given orientation as shown in Fig. \ref{fig:f10}.

\begin{figure}[h!]
  \centering
  \includegraphics[trim={0cm 0.5cm 0cm 0cm}, width=0.8\linewidth]{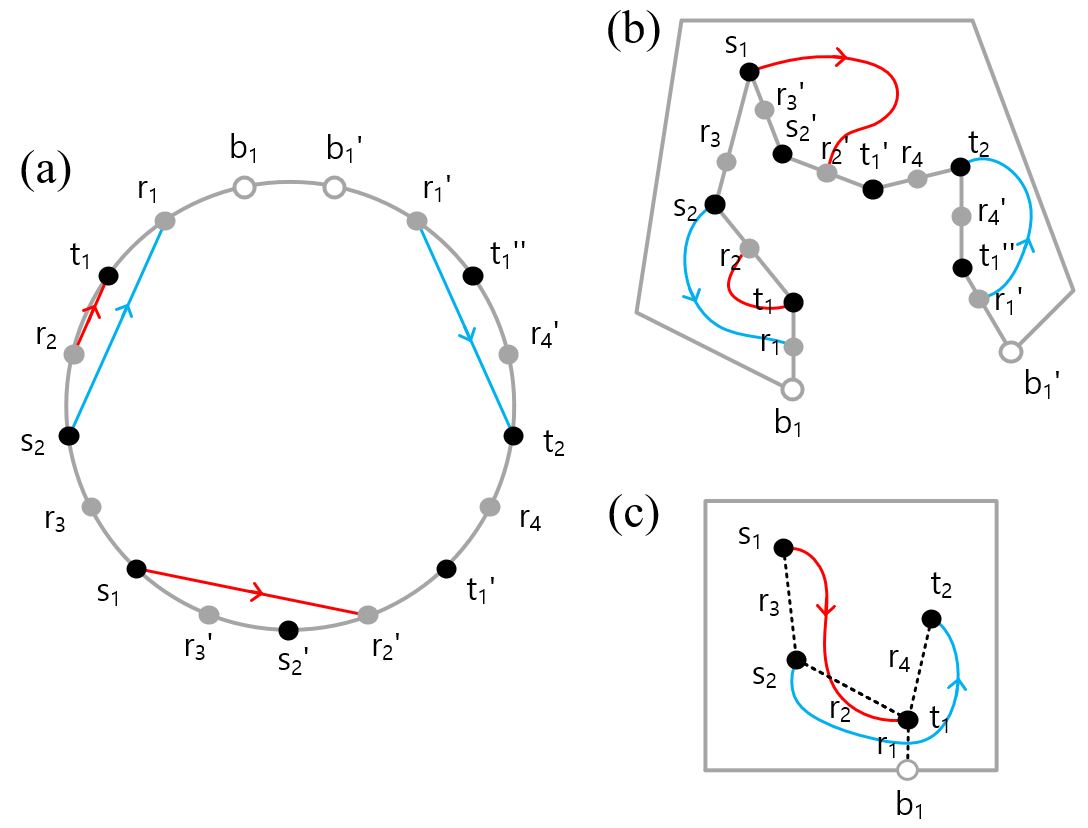}
  \caption{
{\bf Routing Representation in the Circular Frame.} (a) Paths connecting start and end points in the Circular Frame via the point pairs $(r_i, r_i')$ (b) are combined by glueing together $r_i$ with $r_i'$ to form (c) the original substrate layer with the complete routing solution. 
  \label{fig:f11}}
\end{figure}

The Circular Frame is topologically equivalent to the original substrate layer
 where the routing is taking place. The advantage of using the Circular Frame representation of the routing problem is that paths connecting pairs of points are represented as straight line segments connecting points on the boundary of the Circular Frame. These points are either start or end points of the original path, points representing $r_i$ or $r_{i}'$, or points on the original boundary $B$. When a path is connected to $r_i$ or $r_{i}'$ in the Circular Frame, it corresponds in the substrate layer to a path that crosses the associated tree edge $r_i$ as illustrated in Fig. \ref{fig:f11}. A further advantage of the Circular Frame is that line intersections can be easily detected by going through the ordering of line ends on the boundary of the Circular Frame.

The fact that the topological transformation is reversible enables us to solve the routing problem in the simpler Circular Frame environment and then transform the routing solution back to the original substrate layer environment. This is done by reversing the transformation as illustrated in Fig. \ref{fig:f11}. Within the Circular Frame, the routing problem is simply a problem of connecting points on the boundary of a circle with non-intersecting straight line segments as illustrated in Fig. \ref{fig:f11} (a).

\section{Routing Method \label{sec:routingmethod}}

In this section, we outline a method of connecting the nets in the Circular Frame. 
As noted in the section above, although the Circular Frame is topologically equivalent to the original planar substrate layer bounded by $B$, it simplifies the routing problem to a problem of connecting points on a circle with straight line segments that do not intersect in the interior of the circle. 
The following section outlines how the Circular Frame simplifies the routing problem.

Starting from a Circular Frame with no points connected, as illustrated in Fig. \ref{fig:f10} (b), we can choose to connect the first net, \textit{i.e.} $s_1$ with $t_1$. 
Due to the cutting process of the original routing plane, as shown in Fig. \ref{fig:f09}, the end point $t_1$ is split into 3 copies in the Circular Frame, \textit{i.e.} $t_1$, $t_{1}'$ and $t_{i}''$.
We note that in the Circular Frame, connecting $s_1$ to either $t_1$, $t_{1}'$ or $t_{1}''$ is possible.
In the actual routing plane, the choice will determine in which direction the connecting path is going to enter the end point $t_1$ in the original substrate layer environment.

\begin{figure}[h!]
  \centering
  \includegraphics[trim={0cm 0.5cm 0cm 0cm}, width=0.8\linewidth]{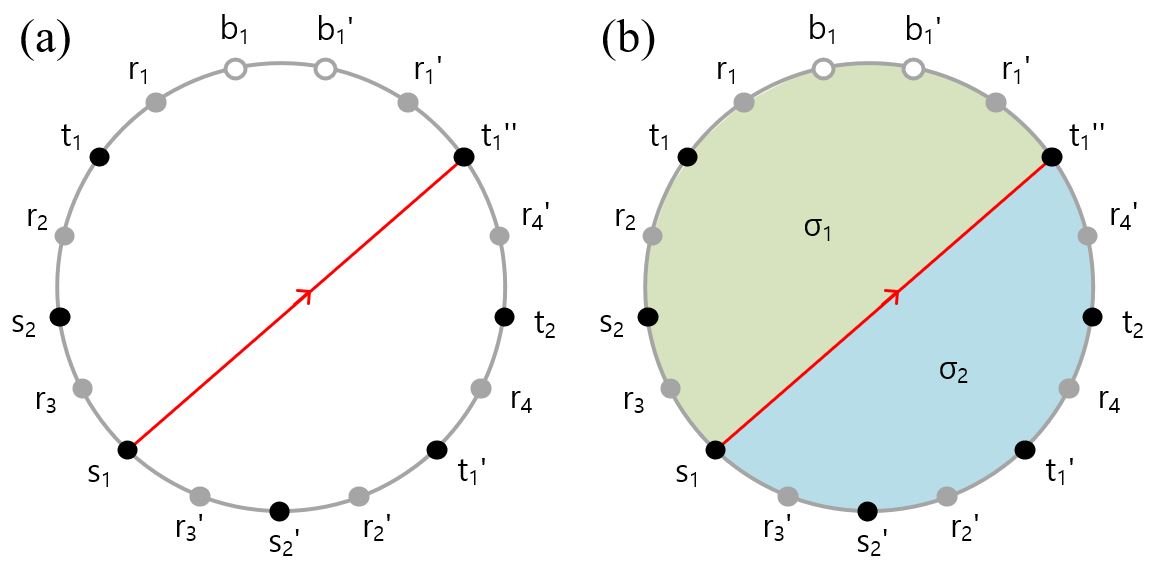}
  \caption{
{\bf Slices in the Circular Frame.} (a) A path connecting two points on the boundary of the Circular Frame (b) divides the Circular Frame into 2 \textit{slices} $\sigma_1$ (green) and $\sigma_2$ (blue). 
  \label{fig:f401}}
\end{figure}

For the moment, without loss of generality, let us assume that we connect in the Circular Frame $s_1$ with $t_1''$ as illustrated in Fig. \ref{fig:f401}. 
Note that any connection between two points in the Circular Frame can be realized in terms of straight line segments that do not intersect in the interior of the Circular Frame. 
Due to the line segment connecting $s_1$ with $t_1''$, the Circular Frame gets divided into two sections, which we call \textit{slices}. 
Fig. \ref{fig:f401} shows the two slices $\sigma_1$ and $\sigma_2$. 
Each slice has its own boundary with a subset of points from the boundary of the Circular Frame. 
For our example in Fig. \ref{fig:f401}, the two slices $\sigma_1$ and $\sigma_2$ have the points $\{s_1, t_1'', r_1', b_1', b_1, r_1, t_1, r_2, s_2, r_3\}$ and $\{s_1, r_3',s_2',r_2',t_1',r_4,t_2,r_4',t_1''\}$ each on their respective boundaries. 
Note that the points that we connected, $s_1$ and $t_1''$, are both shared by the boundary of the two slices. The line segment, which connects $s_1$ with $t_1''$, is precisely the overlap of the two boundaries.

\begin{figure}[h!]
  \centering
  \includegraphics[trim={0cm 0.5cm 0cm 0cm}, width=0.85\linewidth]{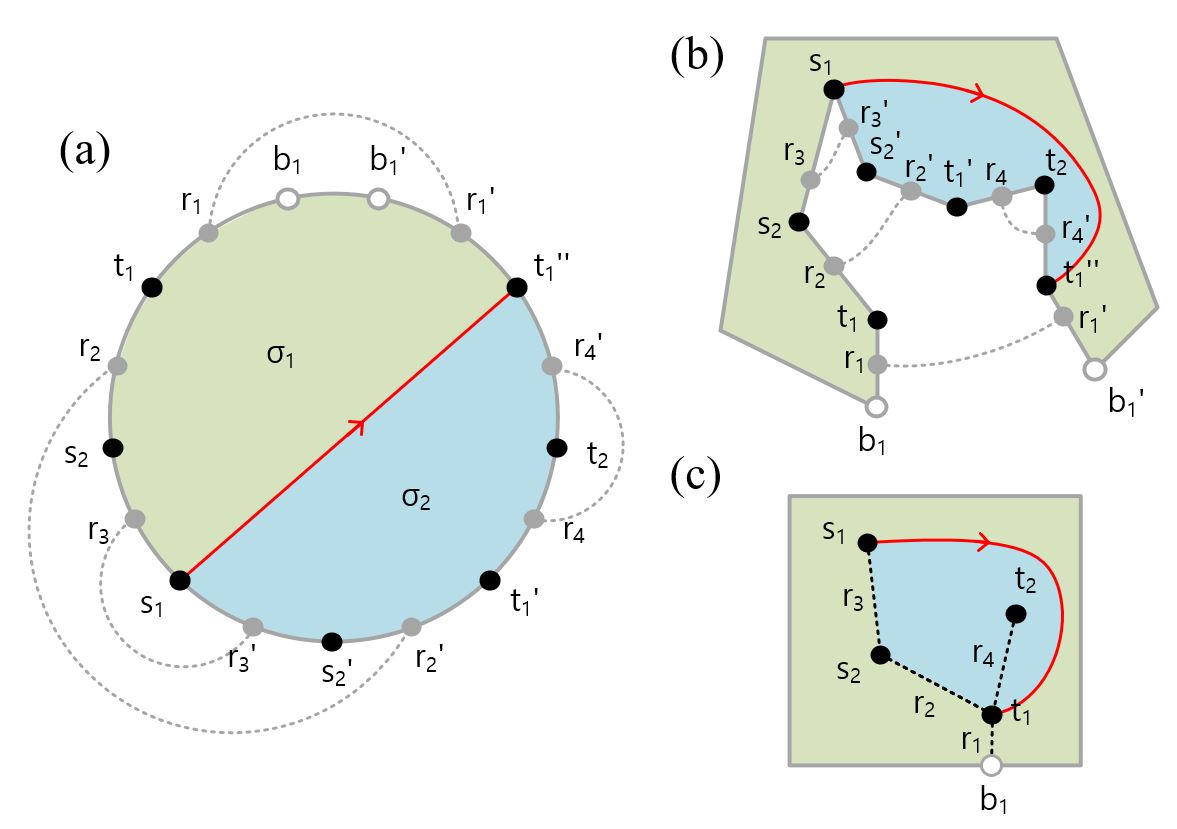}
  \caption{
{\bf Moving Between Slices.} (a) Points $r_i$ and $r_i'$, which always appear in pairs, correspond to the tree edges along which the original substrate layer was cut to give the Circular Frame. (b) The pairs can be pulled together along the dotted lines to give (c) the original substrate layer. We can consider these pairs as `tunnels' along which a connecting path can move between different slices of the Circular Frame.
  \label{fig:f402}}
\end{figure}

As shown in Fig. \ref{fig:f402}, the two slices $\sigma_1$ and $\sigma_2$ are not completely disconnected. 
We recall that the points $r_i$ and $r_i'$ that represent tree edges in the Circular Frame always come in pairs as explained in Section \ref{sec:circularframe}. 
$r_i$ and $r_i'$ precisely identify the tree edges along which the original substrate layer was cut in order to obtain the Circular Frame as illustrated in Fig. \ref{fig:f08}.
Accordingly, they represent points that need to be pairwise glued together when the Circular Frame is transformed back to the original substrate layer environment. 
Fig. \ref{fig:f402} shows these pairwise connections as dotted lines.
The two slices $\sigma_1$ and $\sigma_2$ in Fig. \ref{fig:f402} are connected by the pairs $(r_2, r_2')$ and $(r_3, r_3')$. 
 
\begin{figure}[h!]
  \centering
  \includegraphics[trim={0cm 0.5cm 0cm 0cm}, width=0.85\linewidth]{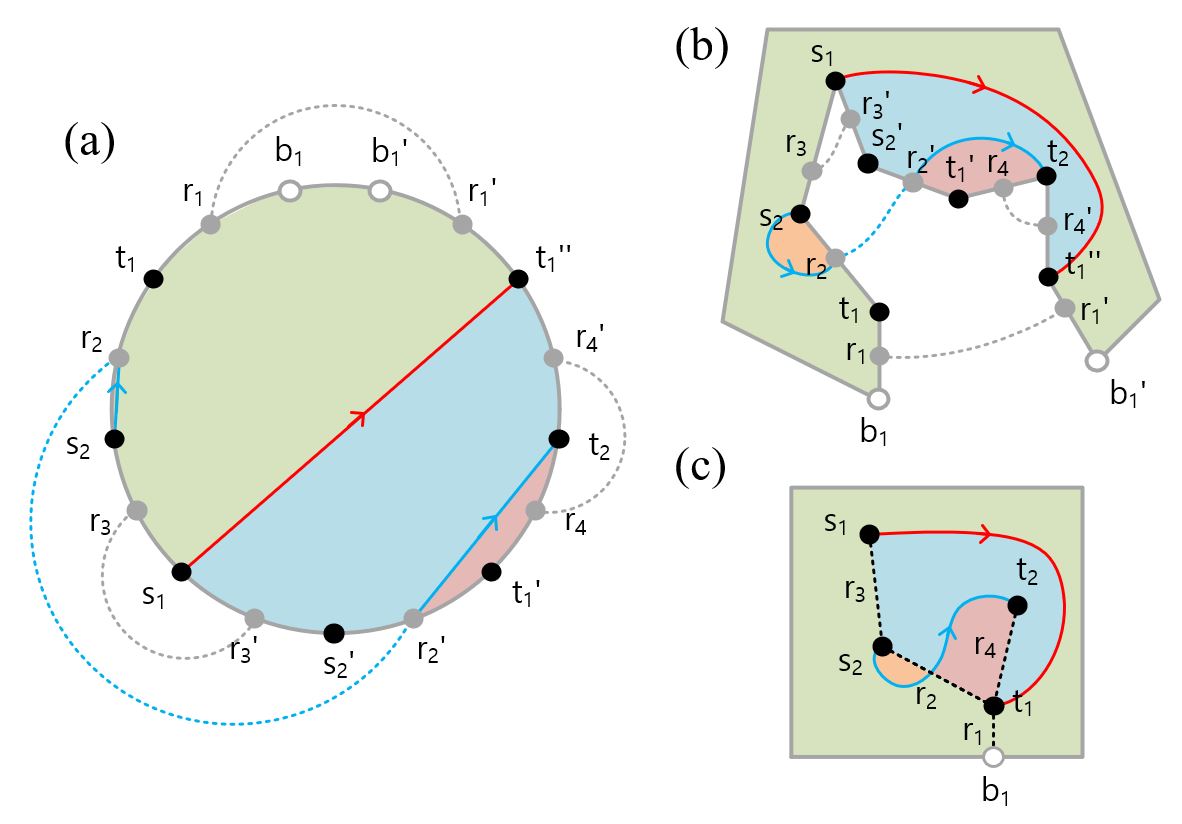}
  \caption{
{\bf Multiple Slices.} (a) By connecting $s_2$ to $t_2$ through  $(r_2, r_2')$, the original slices $\sigma_1$ and $\sigma_2$ are each divided into two slices giving a total of four slices. (b) By glueing together $r_i$ with $r_i'$, we obtain (c) the original substrate layer.
  \label{fig:f403}}
\end{figure}

When we now attempt to connect start point $s_2$, which is on the boundary of $\sigma_1$, with its corresponding end point $t_2$, which is on the boundary of $\sigma_2$, we have to move between the two slices $\sigma_1$ and $\sigma_2$. 
As we noted above, the two slices are connected by the point pairs $(r_2, r_2')$ and $(r_3, r_3')$.
Without loss of generality, by choosing point pair $(r_2, r_2')$, $s_2$ is connected with $r_2$ in $\sigma_1$, and then its partner $r_2'$ is connected with $t_2$ in $\sigma_2$ as illustrated in Fig. \ref{fig:f403}.
Note that by connecting $s_2$ to $t_2$ through the point pair $(r_2, r_2')$, the original slices $\sigma_1$ and $\sigma_2$ are each divided into two slices by the two line segments connecting  $s_2$ with $r_2$ and $r_2'$ with $t_2$. As a result, we end up with a total of four slices.

\begin{figure}[h!]
  \centering
  \includegraphics[trim={0cm 0.5cm 0cm 0cm}, width=0.85\linewidth]{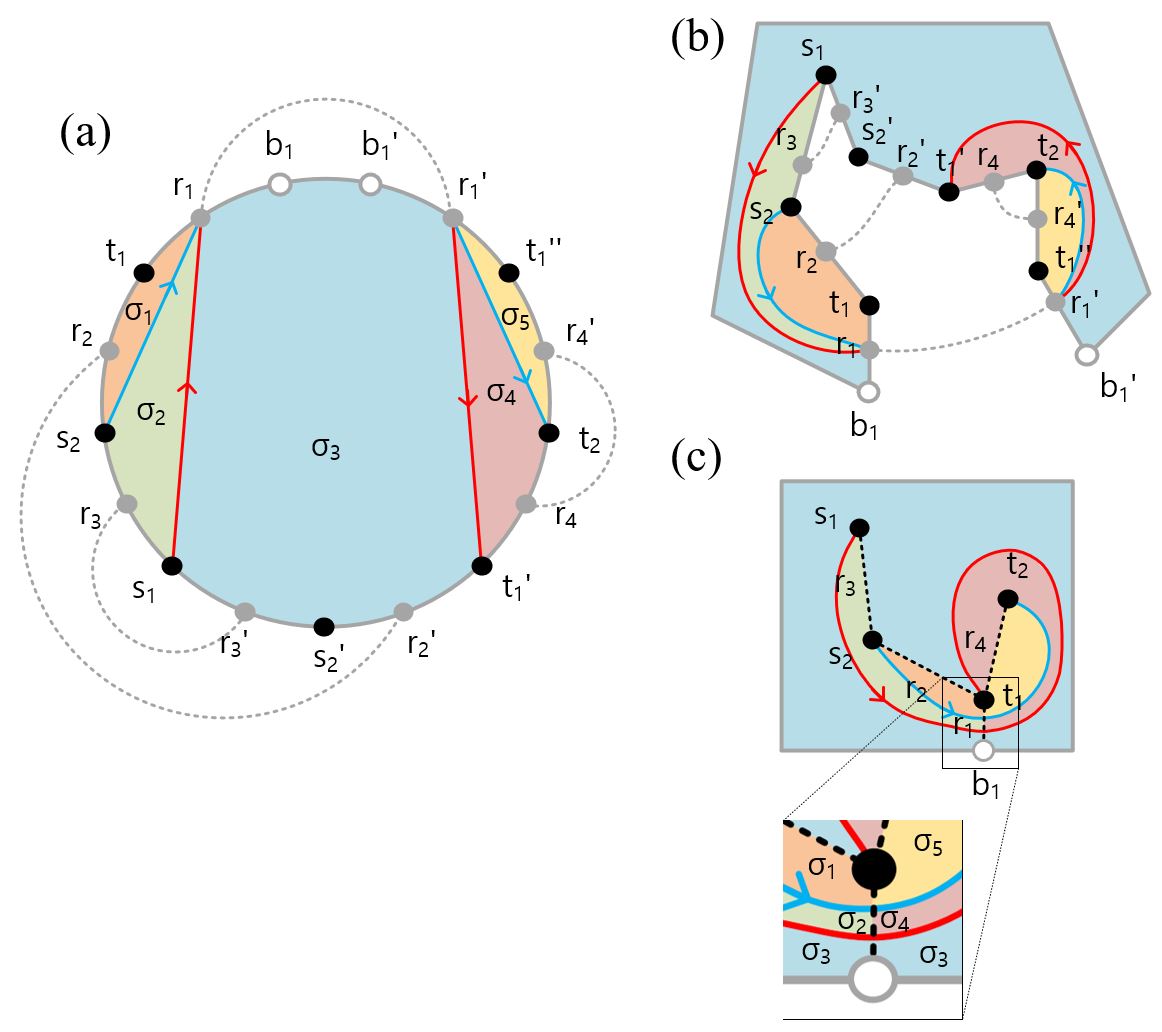}
  \caption{
{\bf Slice Ordering.} (a) Both $(s_1, t_1')$ and $(s_2, t_2)$ are connected through the point pair $(r_1, r_1')$. (b) In order to make sure that when the points $r_1$ and $r_1'$ are glued together the correct slices recombine with each other, (c) we need  $o(\sigma_1, r_1)=o(\sigma_5, r_1')$, $o(\sigma_2, r_1)=o(\sigma_4, r_1')$ and  $o(\sigma_3, r_1)=o(\sigma_3, r_1')$.
  \label{fig:f404}}
\end{figure}

There is also the possibility that more than one path goes through a point pair $(r_1, r_1')$ as shown in Fig. \ref{fig:f404}.
In the example in Fig. \ref{fig:f404}, both $(s_1, t_1')$ and $(s_2, t_2)$ are connected through the point pair $(r_1, r_1')$.  
In such a situation, one has to make sure that the slice containing the origin point and the slice containing the destination point are in the same \textit{order}.
Let us define the order $o(\sigma, r_i)$ of $\sigma$ with respect to the point $r_i$ as the segment number of $\sigma$ attached to $r_i$ in the Circular Frame when one counts anti-clockwise around $r_i$ starting from the boundary of the Circular Frame. In analogy, let us define the order $o(\sigma, r_i')$ of $\sigma$ with respect to the point $r_i'$ as the segment number of $\sigma$ attached to $r_i'$ in the Circular Frame when one counts clockwise around $r_i$ starting from the boundary of the Circular Frame.  
For example, in Fig. \ref{fig:f404}, we note that $o(\sigma_1, r_1)=o(\sigma_5, r_1') = 1$, $o(\sigma_2, r_1)=o(\sigma_4, r_1') = 2$ and $o(\sigma_3, r_1)=o(\sigma_3, r_1') = 3$. 
Accordingly, anything starting in slice $\sigma_1$ can go through $(r_1,r_1')$ to slice $\sigma_5$, not any other slice. 
Similarly, we have $o(\sigma_2, r_1)=o(\sigma_4, r_1')$ and  $o(\sigma_3, r_1)=o(\sigma_3, r_1')$, meaning that anything in slice $\sigma_2$ can be connected to $\sigma_4$ and anything in slice $\sigma_3$ can be connected to $\sigma_4$ via the edge point pair $(r_i, r_i')$.  
Note that slice ordering is essential to make sure that when the edge points are glued together, the correct slices recombine with each other to give the original substrate layer as shown in Fig. \ref{fig:f404} (c).

\begin{figure}[h!]
\centering
\fbox{
\begin{varwidth}{\dimexpr\linewidth-2\fboxsep-2\fboxrule\relax}
\small
\begin{algorithmic}[1]
\For{each $s_i \in S$}
\State select a copy $s_i^{(u)}$ of $s_i$
\State select a copy $t_i^{(v)}$ of $t_i$
\State identify slice $\sigma$ containing $s_i^{(u)}$
\State identify slice $\rho$ containing $t_i^{(v)}$
\If{$\sigma=\rho$}
\State connect $s_i^{(u)}$ and $t_i^{(v)}$
\Else
\State $p=s_i^{(u)}$
\Repeat
\State identify slice $\pi$ containing $p$
\State identify closest $r_k$ to $p$
\State connect $p$ with $r_k$
\State identify slice $\tau$ containing $r_k'$ with $o(\pi, r_k)=o(\tau, r_k')$
\If{$\tau = \rho$}
\State connect $r_k'$ with $t_i^{(v)}$
\Else
\State $p = r_k'$
\EndIf
\Until{$s_i^{(u)}$ is connected with $t_i^{(v)}$}
\EndIf
\EndFor
\end{algorithmic}
\end{varwidth}%
}
\caption{
{\bf A Basic Circular Frame Routing Algorithm.} Pseudocode for connecting nets in the Circular Frame. \label{fig:f_routingmethod}}
\end{figure}

\begin{figure}[h!]
  \centering
  \includegraphics[trim={0cm 0.25cm 0cm 0.5cm}, width=1\linewidth]{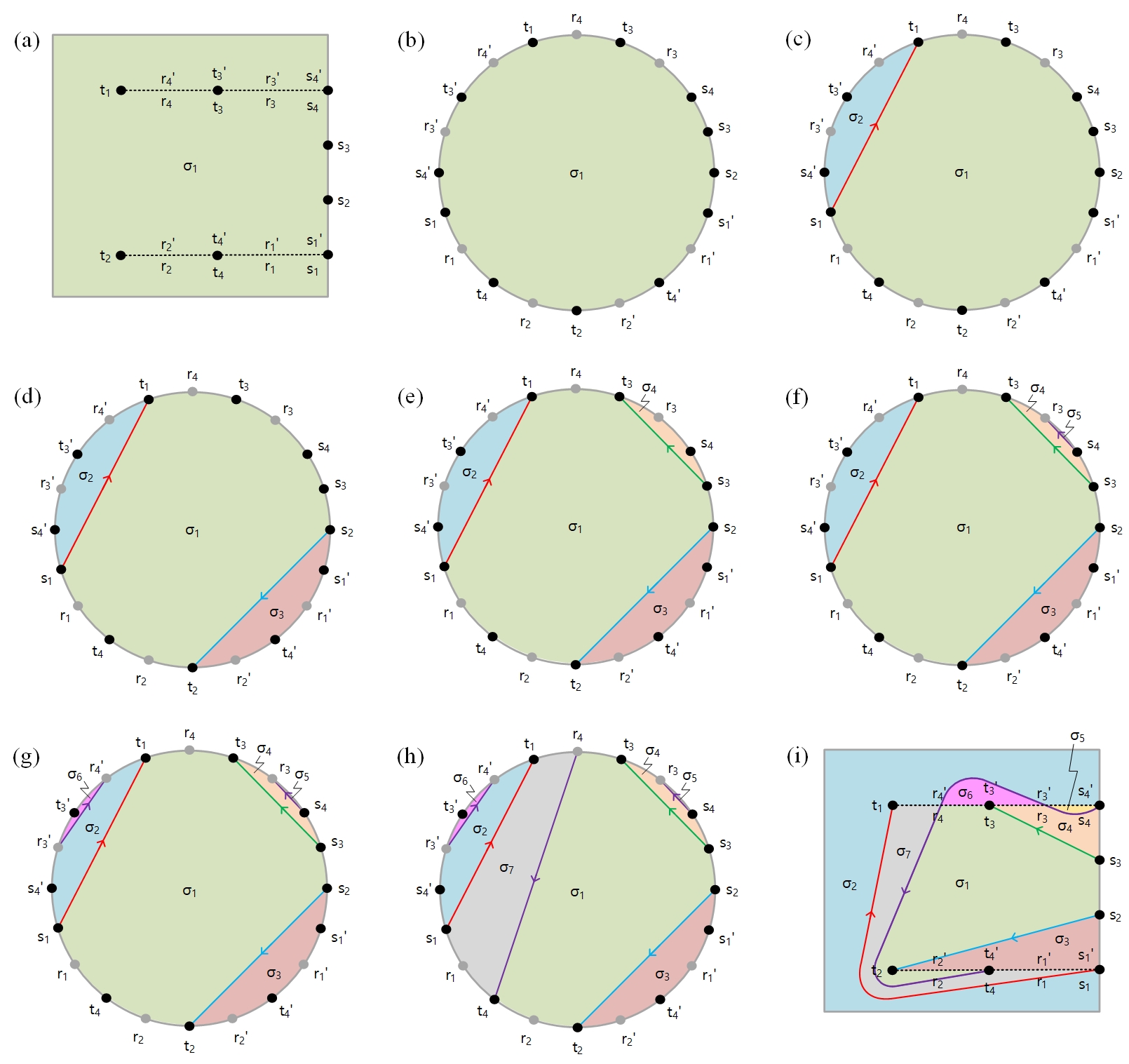}
  \caption{
{\bf Routing Algorithm Example.} (a) The original substrate layer consists of 4 start points $s_i$ placed on the boundary $B$ of the plane and 4 corresponding end points $t_i$. By cutting the substrate layer along the tree edges $r_i$, we obtain (b) the corresponding Circular Frame. 
We select $s_1$ and $t_1$ as the first pair to be connected. (c) Since both $s_1$ and $t_1$ are in slice $\sigma_1$, we connect them. We select $s_2$ and $t_2$ as the second pair to be connected. (d) Because $s_2$ and $t_2$ are both in slice $\sigma_1$, we connect them. We select $s_3$ and $t_3$ as the next pair to be connected. (e) Because $s_3$ and $t_3$ are both in slice $\sigma_1$, we connect them. The final pair to be connected is selected as $s_4$ and $t_4$. Here, $s_4$ is in $\sigma_4$ and $t_4$ is in $\sigma_1$. (f) In $\sigma_4$, the closest tunnelling point to $\sigma_4$ is $r_3$ and we connect $\sigma_4$ with $r_3$. The partner of $r_3$, $r_3'$, is in $\sigma_2$. (g) Since in $\sigma_2$, we still have not $t_4$, we look for the closest tunnelling point to $r_3'$. We identify $r_4'$ and connect $r_3'$ with $r_4'$. (h) The partner of $r_4'$, $r_4$, is in slice $\sigma_1$ where we also have our destination point $t_4$. We connect $r_4$ with $t_4$ and hence have connected using tunnelling points $s_4$ with $t_4$. (i) We reverse transform the Circular Frame with the routing result back to the original substrate layer by glueing together the edge point pairs $(r_i, r_i')$. 
  \label{fig:f405}}
\end{figure}

Following these rules on connecting points in the Circular Frame, we outline a basic algorithm for connecting all points in Fig. \ref{fig:f_routingmethod}. 
In line 12 of the algorithm in Fig. \ref{fig:f_routingmethod}, the closest $r_k$ to a given point $p$ in a slice $\pi$ is identified by the smallest number of points one needs to pass in order to go from $p$ to $r_k$ along the boundary of $\pi$.

We note that the algorithm in Fig. \ref{fig:f_routingmethod} is one example amongst many possible connection algorithms that one can formulate with the help of the Circular Frame. We plan to present variations of this algorithm in future work.
For now, the algorithm in Fig. \ref{fig:f_routingmethod} does not have the aim to find the shortest possible paths between start and end point pairs. Instead, the algorithm in Fig. \ref{fig:f_routingmethod} simply has the aim to achieve full connection for all start and end point pairs.
In fact, with the Circular Frame and the algorithm in Fig. \ref{fig:f_routingmethod}, complete connection is always guaranteed. This is because the Circular Frame is only encoding the topology of the routing problem as illustrated in Fig. \ref{fig:f02} and as a result there is no problem of clearance as it is the case for geometrical routers. 
Furthermore, in signal routing, paths always connect a single start point $s_i$ with a single end point $t_i$ and hence there is no possibility that a path completely encircles points that need to be connected by other paths in the Circular Frame.\footnote{Note that there is an optimization problem in terms of the choice of geometric sketch one uses to represent the routing solution given by its corresponding topological class. The problem of optimization will be the subject of future work.}
Fig. \ref{fig:f405} illustrates an example where the algorithm in Fig. \ref{fig:f_routingmethod} is applied to solve the connection problem in the Circular Frame.

\section{Embedding \label{sec:embedding}}

We call the process of transforming the routing result in the Circular Frame back to the original substrate layer the \textit{embedding} process.
As discussed in the sections above, the Circular Frame can be transformed back to the original substrate layer by glueing together the point pairs $(r_i, r_i')$ for all $i$.
Under this reverse transformation, the topology of the routing result, \textit{i.e.} the identified paths connecting start points with corresponding end points, is preserved. 

The information about which points correspond to which slices in the Circular Frame and the information about which slices are adjacent to each other is called the \textit{topological class} $T$ \cite{fulton} of the routing solution. 
The \textit{geometric sketch} \cite{Dai, fulton} of a topological class is a specific embedding of the connecting paths in the routing solution. 

An example of such a topological class $T_i(P_i,W_i,H_i)$ for a path $\rho_i$ is the following information:
\begin{itemize}
\item The set of \textit{points} $P_i = \{p_k^{(i)}\}$ a path $\rho_i$ passes starting from $s_i$ and ending at $t_i$.
\item The set of \textit{orientations} $W_i = \{w_k^{(i)}\}$ a path $\rho_i$ takes when it passes points $\{p_k^{(i)}\}$. If $\rho_i$ passes $p_k^{(i)}$ anti-clockwise $w_k^{(i)}=+1$, if $\rho_i$ passes $p_k^{(i)}$ clockwise $w_k^{(i)}=-1$, and if $p_k^{(i)} = s_i$ or $t_i$ then $w_k^{(i)}=0$.
\item The set of \textit{heights} $H_i = \{h_k^{(i)}\}$ a path $\rho_i$ has when it passes  points $\{p_k^{(i)}\}$. A height $h_k^{(i)} = m$ indicates that between $\rho_i$ and $p_k^{(i)}$ there are $m-1$ other paths passing $p_k^{(i)}$. If $p_k^{(i)} = s_i$ or $t_i$ then $h_k^{(i)}=0$.
\end{itemize}
Note that $T_i(P_i,W_i,H_i)$ for any path $\rho_i$ can be obtained from the Circular Frame of the routing solution.\footnote{
We note that one can introduce several other topological classes that encapsulate the routing result in the Circular Frame, for instance including the edges $r_i$. We hope to present further versions in future work.} 
In the example in Fig. \ref{fig:f405}, for path $\rho_4$ connecting $s_4$ with $t_4$, $T_4$ is given by $P_4=(s_4, t_3, t_2, t_4)$, $W_4=(0,+1,+1,0)$ and $H_4=(0,1,1,0)$.

\begin{figure}[h!]
  \centering
  \includegraphics[trim={0cm 0cm 0cm 0cm}, width=0.65\linewidth]{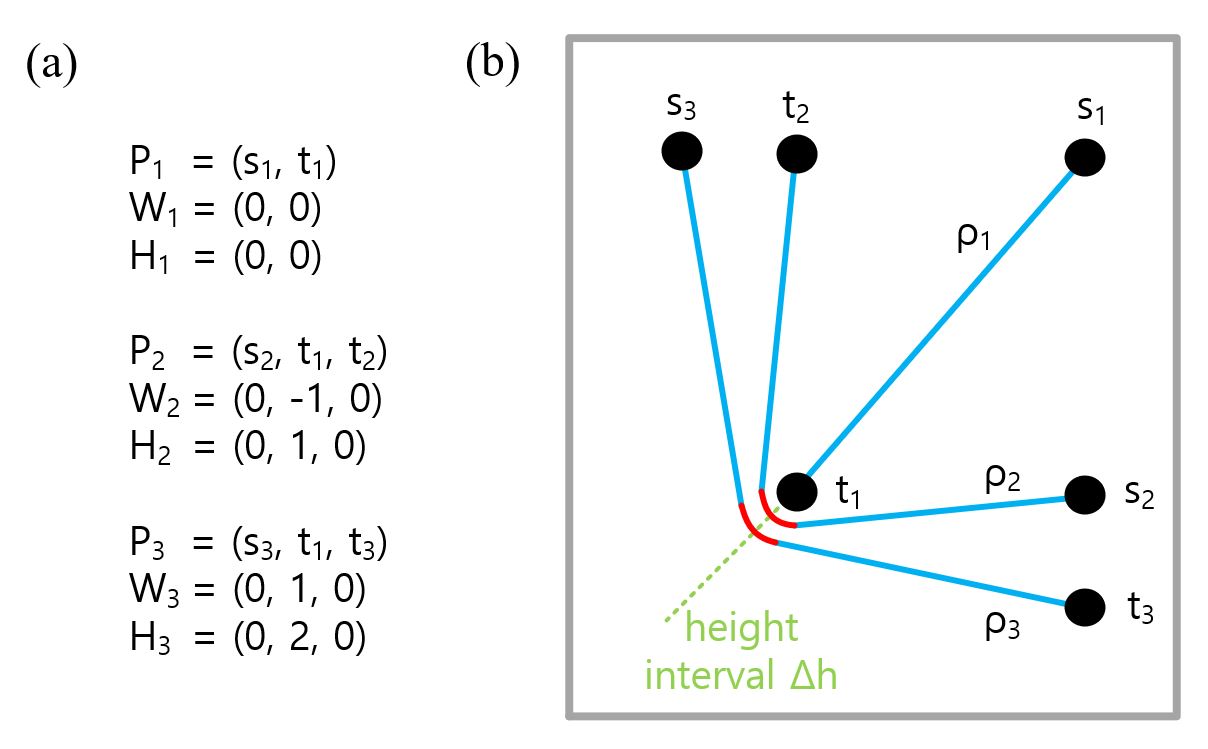}
  \caption{
{\bf Topological Class and Rubber-Band Sketch.} (a) A topological class $T(P,W,H)$ for a 3-path routing solution with (b) the corresponding rubber-band sketch representation. Note that the separation between paths passing $t_1$ is given by the height interval $\Delta h = 0.5$. The paths are made of line segments (blue) and concentric arcs (red).
  \label{fig:f16B}}
\end{figure}

We make use of the \textit{rubber-band sketch} from \cite{Dai, Dai_1991} in order to represent the topological class of the routing result from the Circular Frame on the original planar environment.
Fig. \ref{fig:f03} (a) shows the rubber-band sketch of the same topological class represented in Fig. \ref{fig:f03} (b).
A characteristic feature of the rubber-band sketch is that paths are represented as line segments that can have any angle and the line segments are connected by arcs whenever the path passes a point.
Fig. \ref{fig:f16B} illustrates an example of a topological class and its corresponding rubber-band representation. 

For the purpose of this work, which is to present a new topological routing method that results in a topological class of a fully-connected routing result via the Circular Frame, we keep the review on topological classes and the rubber-band sketch representation short and refer to the works in \cite{Dai, Dai_1991}.

\section{Experiment \label{sec:experiment}}

Let us design experiments to compare the performance of the proposed routing algorithm based on the Circular Frame (CF) with variations of the A*-algorithm (AS). 

\subsection{General Setup}

Let us first construct a planar environment with a boundary $B$ having $-50.0\leq x \leq 50.0$ and $-50.0 \leq y \leq 50.0$.
The start points $S=\{s_1,\dots,s_n\}$ and end points $T=\{t_1,\dots,t_n\}$ are represented as circles with radius $r=0.5$ and their positions are given by the coordinates of their centers.
For our experiment, we vary $n$ by setting it to $n=2,4,6,8,10$.\footnote{Note that in many substrate designs, nets can be grouped into independent substrate segments containing on average around 10 start and end points in signal routing.} 
The number of start points $S$ increases by adding consecutively $(50.0,\pm 4.0)$, $(50.0,\pm 12.0)$, $(50.0,\pm 20.0)$, $(50.0,\pm 28.0)$ and $(50.0,\pm 36.0)$.

\begin{figure}[h!]
  \centering
  \includegraphics[trim={0cm 0.8cm 0cm 0.7cm}, width=0.7\linewidth]{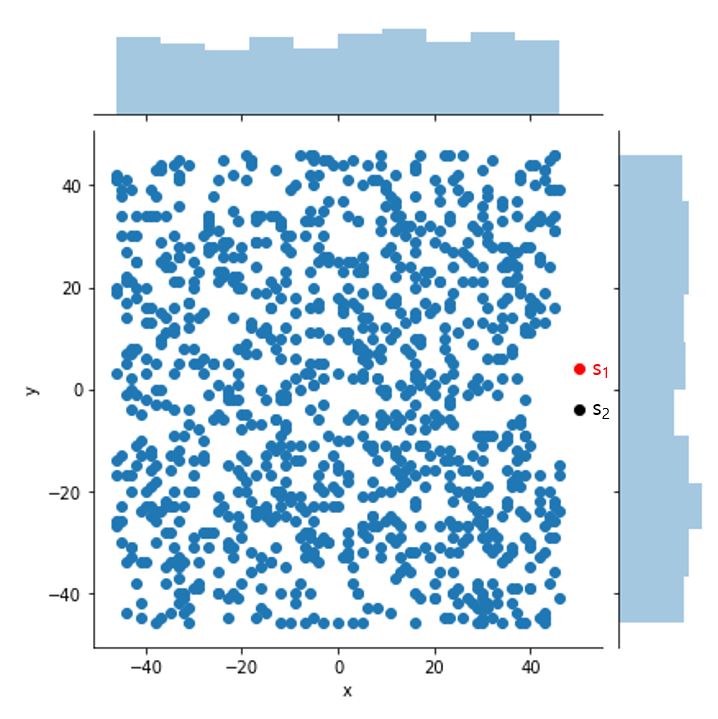}
  \caption{
{\bf Random Positions for End Points.} $N=1000$ randomly generated end points $t_1$ for $s_1$ in an environment with $n=2$ start and end point pairs. We have shown all end points $t_1$ for $s_1$ (red) at once to illustrate the random distribution of the points within the boundary of the environment.
  \label{fig:f601}}
\end{figure}

For each $n$, we generate $N=1000$ end ball sets $T$ whose coordinates are generated randomly within the boundary of the planar environment.
The randomly generated end points $t_i$ have a minimum center-to-center separation $d_{\min}(t_i,t_j)=11$ to other end points $t_j$ as well as a minimum center-to-center separation $d_{\min}(t_i,s_j)=11$ to start points $s_j$.
The randomly generated end points also satisfy a minimum distance $d_{\min}(t_i,b_j)=3$ to any boundary point $b_j\in B$ of the plane.

We call each generated set $(S,T)$ a routing environment $E_{h=1\dots N}$.
Fig. \ref{fig:f601} shows an environment with $n=2$, where all $N=1000$ randomly generated end points for $s_1$ are illustrated simultaneously in order to illustrate that the randomly generated points are evenly distributed on the bounded plane.

\subsection{Measurements}

For each environment $E_h$, the routing problem is to connect  all $s_i$ with the corresponding $t_i$ with non-intersecting paths. 
We run different routing algorithms for each environment $E_h$ and measure the time $t_{h}$ that the algorithms take to complete the routing for all nets.
Note that all algorithms are run on a laptop with CPU at 1.80 GHz (Intel i7-8550U) and 8 GB memory.
If any of the nets are left disconnected, we label the routing result as incomplete. 
 
For completed routing environments, we also measure for each connecting path between $s_i$ and $t_i$ the Euclidean path length $l_i^h$. The mean path length $\overline{l}^h$ and the corresponding standard deviation $\sigma^h$ for all connecting paths in $E_h$ are also obtained. 

In addition, we also measure the \textit{Manhattan distance}  between the start node $s_i$ and corresponding end node $t_i$, 
\begin{equation}
d_{i}^h(s_i, t_i) = |x(s_i) - x(t_i)| + |y(s_i) - y(t_i)| ~,~  
\end{equation}
and compare it to the Euclidean path length $l_i^h$ of the path that was found by the chosen routing algorithm. In particular, we calculate the ratio $r_i^h(s_i, t_i)=l_i^h(s_i, t_i)/d_{i}^h(s_i, t_i)$. 
The Manhattan distance is the shortest path length between start and end points on a square grid and is a measure of how direct a path has been taken between a start point and its corresponding end point. Accordingly, a smaller $r_i^h$ indicates that the path is closer to the shortest path on a square grid.

For all our measurements, we have two different types of means. A measurement $X_i^h$ corresponding to $(s_i, t_i)$ in environment $E_h$ can be averaged over all paths in $E_h$ to give $\overline{X}^h=\frac{1}{n} \sum_i X_i^h$ and then further averaged over all environments $E_h$ to give $\overline{X}=\frac{1}{N}\sum_h \overline{X}^h$. The corresponding standard deviation of sample means is denoted as $\sigma_{\overline{X}}$. We are going to use this notation when we summarize our experimental results in Section \ref{sec:results}.

\subsection{A*-Algorithm}

Let us give a brief overview of the A*-algorithm used in this work for the purpose of benchmarking our new routing algorithm based on the Circular Frame. The reader is referred to previous work for a more extended overview of the A*-algorithm \cite{hart1968formal}.

\begin{figure}[h!]
  \centering
  \includegraphics[trim={0cm 0cm 0cm 0cm}, width=0.7\linewidth]{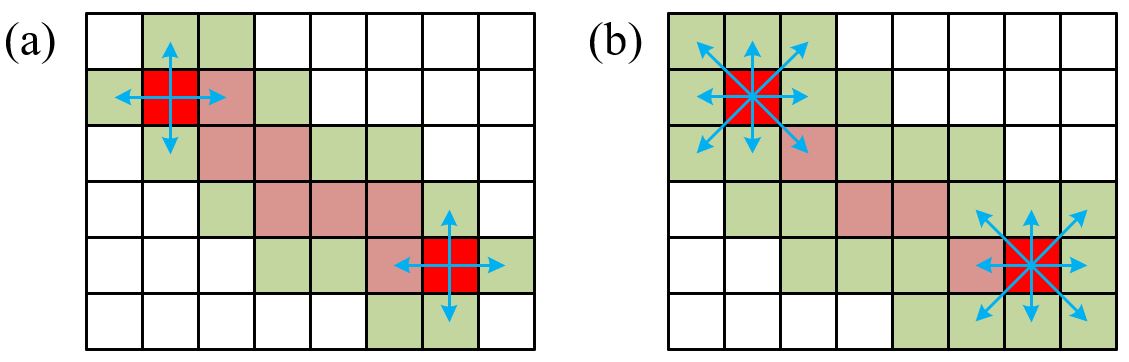}
  \caption{
{\bf A*-Algorithm.} Implementations of the A*-algorithm using a square grid with (a) 4 directions of movement and a Manhattan distance heuristic (AS1), and (b) 8 directions of movement and a Chebyshev distance heuristic (AS2).
  \label{fig:f602}}
\end{figure}

The A*-algorithm is a graph traverser algorithm, which at each iteration of the algorithm extends a tree of candidate paths originating from the start node $s_i$ until one of the branches of the tree reaches the end node $t_i$. 
The incremental extension is made at a given node $p$ of the graph if a cost function $f(p)$ is minimized by the extension. 
The cost function is defined as $f(p) = g(p) + h(p)$,
where $g(p)$ is the cost of the path from the start node to $p$ and $h(p)$ is the \textit{heuristic} function that estimates the cost of the cheapest path from $p$ to the end node. Without loss of generality we define $g(p)$ as the Euclidean path length from the start point to $p$ unless the path intersects with another path in which case its value is set to infinity.

\begin{table}[h!]
  \centering
  \caption{
{\bf Routing Completion Results.}
  \label{table:t01}}
 \begin{tabular}{cccccc}
 \hline\hline
 $n$ & 2 & 4 & 6 & 8 & 10
 \\
 \hline
 $N$ & 1000 & 1000 & 1000 & 1000 & 1000 
 \\
 \hline
 $N_{CF}$ & 1000 & 1000 & 1000 & 1000 & 1000
 \\ 
 $N_{AS1}$ & 1000 & 931 & 742 & 424 & 209
 \\ 
 $N_{AS2}$ & 934 & 913 & 749 & 496 & 220 
 \\
 \hline
 $N_{C}$ & 934 & 868 & 628 & 320 & 116
 \\
 \hline\hline
 \end{tabular}
\end{table}

\begin{figure}[h!]
  \centering
  \includegraphics[trim={0cm 0cm 0cm 0cm}, width=0.7\linewidth]{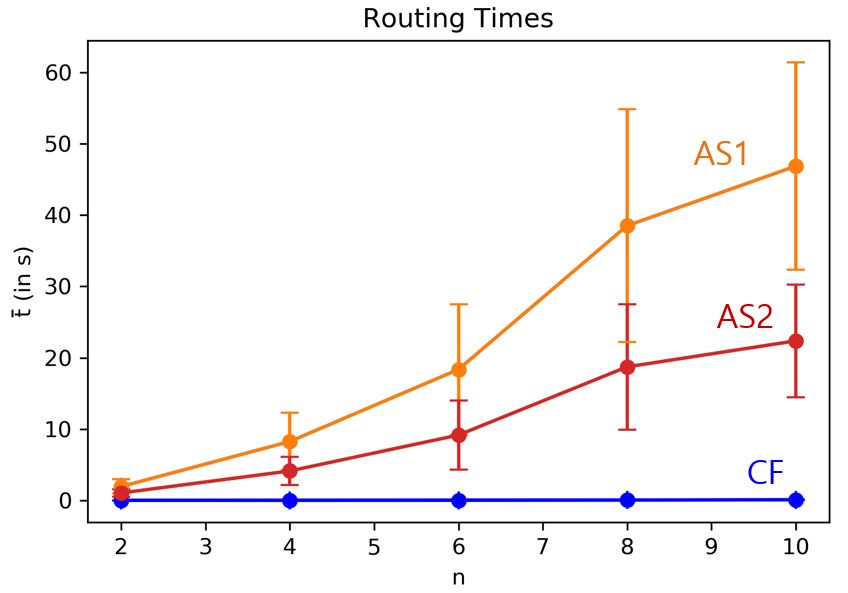}
  \caption{
{\bf Average Routing Times.} Average routing times $\overline{t}$ to complete the routing problem at given $n$ for the Circular Frame algorithm (CF) and A*-algorithms (AS1 and AS2).  The error bars show the standard deviation $\sigma_{\overline{t}}$ of $\overline{t}$. 
\label{fig:f701}
}
\end{figure}

\begin{figure*}[h!]
  \centering
  \includegraphics[trim={0cm 1cm 5cm 0cm}, width=0.9\linewidth]{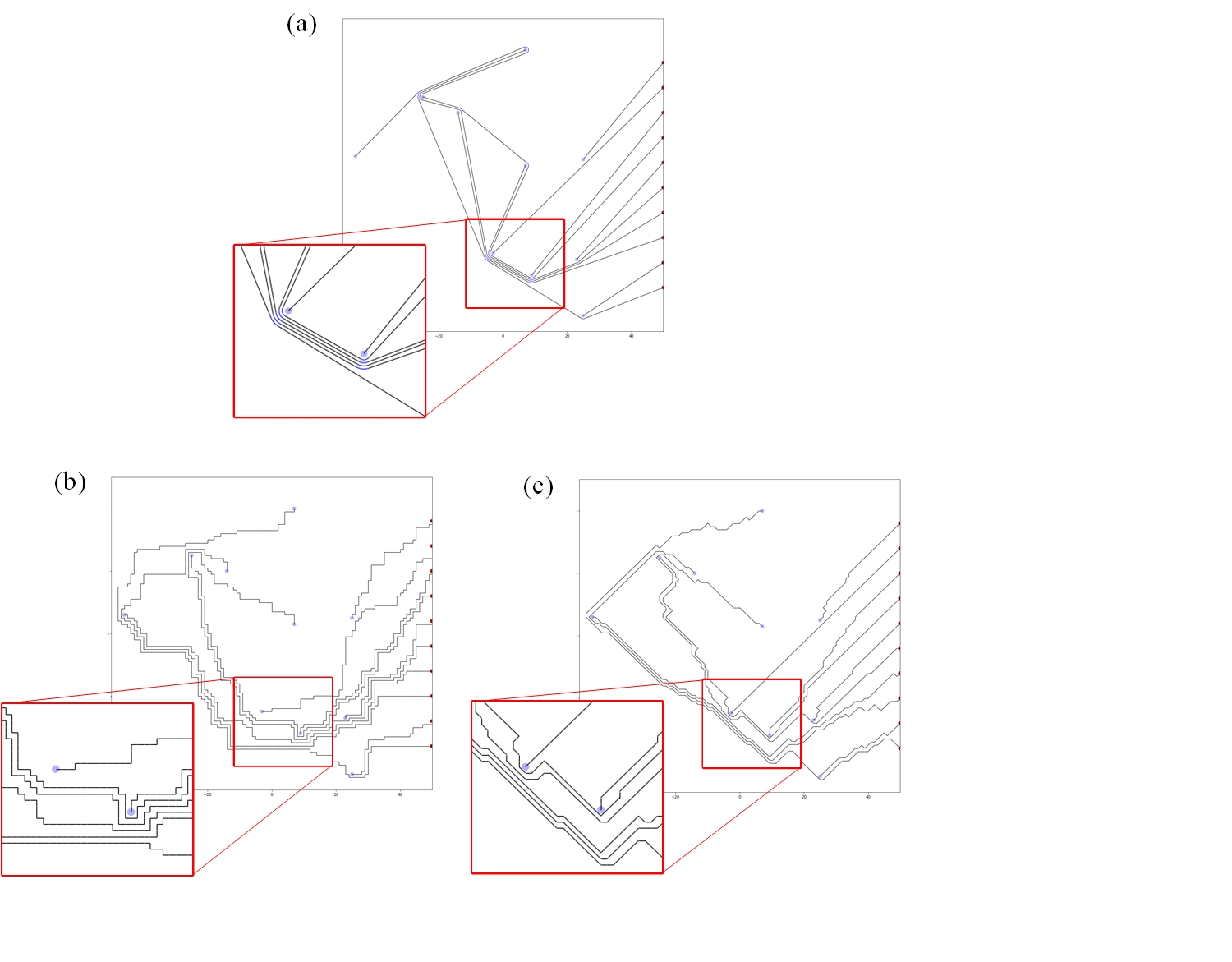}
  \caption{
{\bf Routing Sample.} Complete routing result for the same routing environment under (a) the Circular Frame algorithm, (b) the A*-algorithm with Manhattan distance heuristic (AS1) and (c) the A*-algorithm with the Chebyshev distance heuristic (AS2).
  \label{fig:f702}}
\end{figure*}

For our experiments, we consider two different implementations of the A*-algorithm. 
The first implementation (AS1) uses as the graph the integer square grid of the plane bounded by $B$ with the neighbourhood of a given node defined by the 4 direction vectors $(\pm 1,0)$ and $(0,\pm 1)$. 
The corresponding heuristic uses the Manhattan distance between a given node $p$ and the corresponding destination $t_i$, 
\begin{equation}
h_{AS1}(p)= D (dx + dy) ~,~
\end{equation}
where here we set $D=1$, and 
$dx = |x(p) - x(t_i)|$ and $dy = |y(p) - y(t_i)|$.
The second implementation (AS2) defines the neighbourhood of a node in the integer square grid by the 8 direction vectors $(\pm 1,0)$, $(0,\pm 1)$ and $(\pm 1,\pm 1)$. We use here the \textit{Chebyshev distance} heuristic,
\begin{equation}
h_{AS2}(p) = D_1 (dx + dy) + (D_2 - 2 D_1) \min(dx, dy) ~~,~~
\end{equation}
where we set $D_1 = 1$ and $D_2 = 1$.
Fig. \ref{fig:f602} illustrates the difference between the two implementations of the A*-algorithm that we use in this work.

For our experiments we use Python implementations of the above A*-algorithms and a Python implementation of the Circular Frame algorithm described in Section \ref{sec:routingmethod}.

\section{Results and Discussions \label{sec:results}}

Let us summarize the results of the experiments in the following section.

\subsection{Reliability and Performance}

\begin{table}[h!]
  \centering
  \caption{
{\bf Average Routing Completion Times.}
  \label{table:troutingtimes}}
 \begin{tabular}{ccccccc}
 \hline\hline
 & $n$ & 2 & 4 & 6 & 8 & 10
 \\
 \hline
 & $N_{C}$ & 934 & 868 & 628 & 320 & 116
 \\
 \hline
 \multirow{2}{*}{CF} & $\overline{t}$ & 0.0041 & 0.0122 & 0.0269 & 0.0494 & 0.0846
 \\ 
 & $\sigma_{\overline{t}}$ &0.0009 & 0.0032 & 0.0088 & 0.0163 & 0.0312
 \\
 \hline 
 \multirow{2}{*}{AS1} & $\overline{t}$ & 1.94452 & 8.25649 & 18.35662 & 38.50646 & 46.85451
 \\ 
 & $\sigma_{\overline{t}}$ & 1.0035 & 4.0399 & 9.1618 & 16.3210 & 14.5456
 \\
 \hline
 \multirow{2}{*}{AS2} & $\overline{t}$ & 1.04229 & 4.12858 & 9.16055 & 18.71960 & 22.36284
 \\ 
 & $\sigma_{\overline{t}}$ & 0.4988 & 2.0021 & 4.8156 & 8.7838 & 7.8897
 \\
 \hline\hline
 \end{tabular}
\end{table}

Table \ref{table:t01} shows the number of successfully completed routing problems under the Circular Frame algorithm ($N_CF$), the A*-algorithm under the Manhattan distance heuristic ($N_{AS1}$) and the A*-algorithm under Chebyshev distance heuristic ($N_{AS2}$). Originally $N=1000$ routing environments were generated as outlined in Section \ref{sec:experiment}. We can see that for all number of start points $n$, the Circular Frame algorithm consistently completes the routing for all generated environments, whereas the number of routing failures increases with increasing $n$ for the two implementations of the A*-algorithm.

Table \ref{table:t01} also shows the number $N_C$ of environments where the routing was completed by all 3 tested routing algorithms. For these completed environments and for each $n$, we measure the mean routing times $\overline{t}$ with the corresponding standard deviations $\sigma_{\overline{t}}$.
Fig. \ref{fig:f701} shows that the average routing time for the Circular Frame algorithm stays consistently below the average routing times for the two implementations of the A*-algorithm.

From the routing completion numbers in Table \ref{table:t01} and the average routing times illustrated in Fig. \ref{fig:f701}, we conclude for the test environments that the Circular Frame algorithm is more reliable and faster than the two implementations of the A*-algorithm. This is not a surprising result since the number of points that needs to be traversed on the boundary of the Circular Frame is far less than the grid points used for grid-based geometrical routers. Moreover, as a topological router, the Circular Frame algorithm does not suffer from clearance problems as the A*-algorithms do as illustrated in Fig. \ref{fig:f02}. 

\subsection{Routing Accuracy}

Table \ref{table:t701} shows the grand mean of the path lengths $\overline{l}$ with the corresponding standard deviation of the mean $\sigma_{\overline{l}}$ for the $N_C$ completed routing environments under the 3 tested algorithms.
We observe that paths connecting nets under the Circular Frame algorithms tend to be shorter than for the two implementations of the A*-algorithm for all $n$. 
This is not surprising since the rubber-band sketch representation of the resulting topological class uses arcs and any-angle straight lines for paths, making the overall routing more compact than grid-based representations.

\begin{table*}[h!]
  \centering
  \caption{
{\bf Routing Path Length Results.}
  \label{table:t701}}
 \begin{tabular}{ccccccc}
 \hline\hline
 & $n$ & 2 & 4 & 6 & 8 & 10
 \\
 \hline
 & $N_{C}$ & 934 & 868 & 628 & 320 & 116
 \\
 \hline
 \multirow{2}{*}{Manhattan} & $\overline{d}$ 
 & 73.926 
 & 76.836 
 & 77.789 
 & 78.220 
 & 79.055
 \\ 
 & $\sigma_{\overline{d}}$ 
 & 29.004 
 & 29.090 
 & 30.362 
 & 30.964 
 & 30.755
 \\
 \hline
 \multirow{4}{*}{CF} & $\overline{l}$ 
 & 62.787 
 & 76.221 
 & 85.638 
 & 92.658 
 & 103.115
 \\ 
 & $\sigma_{\overline{l}}$ 
 & 18.662  
 & 19.451 
 & 24.711 
 & 26.031 
 & 34.285
 \\
 & $\overline{r}$
 & 0.951 
 & 1.151 
 & 1.308
 & 1.439
 & 1.591
 \\ 
 & $\sigma_{\overline{r}}$ 
 & 0.386 
 & 0.574 
 & 0.743
 & 0.862 
 & 1.084
 \\
 \hline
 \multirow{4}{*}{AS1} & $\overline{l}$ 
 & 87.565 
 & 115.357
 & 126.014
 & 130.207
 & 128.716
 \\ 
 & $\sigma_{\overline{l}}$ 
 & 32.862
 & 35.974
 & 35.765
 & 32.015
 & 25.660
\\
 & $\overline{r}$
 & 1.327 
 & 1.754 
 & 1.940
 & 2.034
 & 2.005
 \\ 
 & $\sigma_{\overline{r}}$ 
 & 0.663 
 & 0.990
 & 1.180
 & 1.244 
 & 1.301 
 \\
 \hline
 \multirow{4}{*}{AS2} & $\overline{l}$ 
 & 83.903
 & 108.047
 & 116.246
 & 120.616
 & 119.095
 \\ 
 & $\sigma_{\overline{l}}$ 
 & 32.070
 & 31.939
 & 30.476
 & 27.257
 & 20.304
 \\
 & $\overline{r}$
 & 1.277
 & 1.645 
 & 1.791
 & 1.889
 & 1.857
 \\ 
 & $\sigma_{\overline{r}}$ 
 & 0.669 
 & 0.927
 & 1.058
 & 1.153 
 & 1.181 
 \\ 
 \hline\hline
 \end{tabular}
\end{table*}

We also calculated the ratio $r_i^h$ between the Euclidean path length $l_i^h$ and the corresponding Manhattan distance $d_i^h$ between the connected start and end points $(s_i, t_i)$ for all completed $E_h$. The grand mean of the ratio $\overline{r}$ with $\sigma_{\overline{r}}$ is shown in Table \ref{table:t701}.
As noted in Section \ref{sec:experiment}, a smaller ratio $r$ indicates that the connection is closer to the shortest path on a square grid. 
As we can see in Table \ref{table:t701}, the Circular Frame algorithm at $n=2$ has a mean ratio $\overline{r}<1$, indicating that for some routing results, the Circular Frame identified connecting paths that are even shorter than the Manhattan distance $d$. 
Moreover, consistently the Circular Frame algorithm achieved on average a smaller value of $\overline{r}$ than the two implementations of the A*-algorithm, indicating that overall the Circular Frame algorithm found more direct and hence more accurate connections.

\begin{figure}[h!]
  \centering
  \includegraphics[trim={0cm 0.3cm 0cm 0cm}, width=0.95\linewidth]{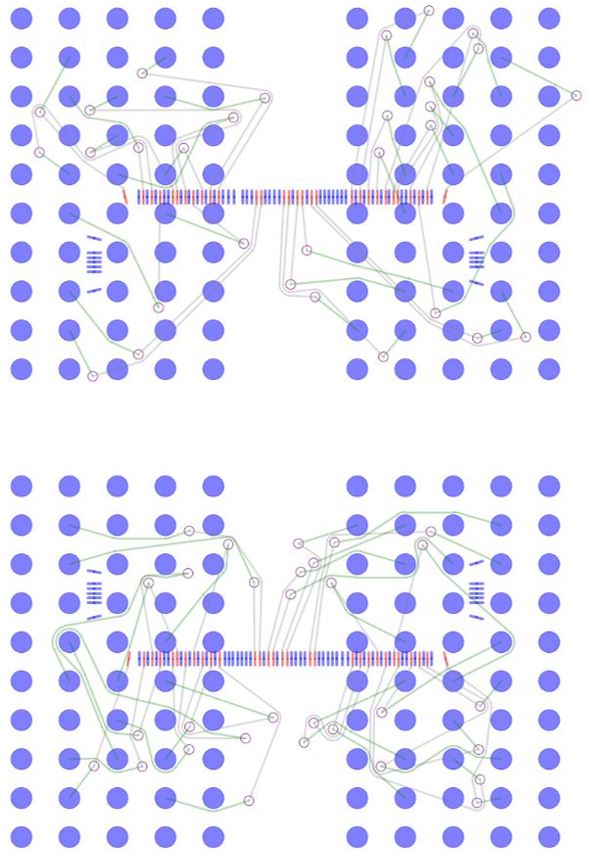}
  \caption{
{\bf FBGA Routing Sample.} 2-layered FBGA package substrate that has been connected using the Circular Frame routing algorithm. Layer 1 consists of fingers (red and blue) and net connections (gray) while layer 2 consists of solder balls (blue) and net connections (green). The two layers are connected by vias (white circles).
  \label{fig:fullrouting}}
\end{figure}

Fig. \ref{fig:f702} shows the completed routing results under the Circular Frame algorithm and the two implementations of the A*-algorithm for a given routing environment. 
We note that our observations here are as expected since the Circular Frame algorithm avoids as a topological router problems caused by a lack of clearance as discussed in Fig. \ref{fig:f02}. Furthermore, the rubber-band sketch representation optimizes the length of the connecting paths in comparison to grid-based geometrical routers.

\section{Conclusions}

In this paper, we have proposed a new method based on our earlier work in \cite{2021arXiv210503386S} for solving substrate routing problems using  topology. 
Our proposed topological transformation of the original routing environment into the Circular Frame has accelerated in experiments the substrate routing process significantly in comparison to grid-based geometrical routers such as the A*-algorithm.
Moreover, the Circular Frame representation guarantees  for substrate routing problems with start and end point pairs full connection as a topological router.

In addition, Fig. \ref{fig:fullrouting} shows a 2-layered substrate for a FBGA package with 200 solder balls and a completed routing design that was obtained using our new Circular Frame algorithm.
Our experiments and the positive routing results on real semiconductor package substrates are a clear indication that our new Circular Frame routing method has the potential to significantly improve and at the end fully automate the package substrate routing process.

We note that the Circular Frame routing algorithm can lead to different routing solutions given by topological classes depending on via placement, spanning tree generation and even net ordering during the routing process in the Circular Frame.  Moreover, in our work we have given only a single basic routing method based on the Circular Frame representation and depending on other routing algorithms based on the Circular Frame representation, the routing result can differ significantly. Finding the most optimal routing solution based on the Circular Frame representation depends on metrics such as wire length or wire widths and is an optimization problem that we hope to cover in future work.

We are currently testing the Circular Frame routing algorithm on larger FBGA packages and other package designs. Moreover, beyond semiconductor package design, we are applying our routing method on problems related to the design of printed circuit boards (PCB) and the logistics and  manufacturing industry. We hope to report on our progress in these areas in future work.

\section*{Acknowledgements}

The authors would like to thank Minsoo Kim for suggesting the problem
and Seungjai Min and Youngjae Gwon at Samsung SDS for helpful discussions.
The authors are also grateful to Joung Oh Yun and Minkyu Jung for helpful guidance during the project, and Chanho Min for collaborating on an earlier project.


\bibliographystyle{eg-alpha-doi} 
\bibliography{mybib2}      

\newcommand{\etalchar}[1]{$^{#1}$}
\begin{thebibliography}{\uppercase{WHJ{\etalchar{*}}20}}

\bibitem[{Alb}01]{920691}
\textsc{{Albrecht} C.}:
\newblock Global routing by new approximation algorithms for multicommodity
  flow.
\newblock \emph{IEEE Transactions on Computer-Aided Design of Integrated
  Circuits and Systems 20}, 5 (2001), 622--632.

\bibitem[AZN17]{8102207}
\textsc{{Ali} A., {Zeeshan} M., {Naveed} A.}:
\newblock A network flow approach for simultaneous escape routing in pcb.
\newblock In \emph{2017 14th International Conference on Smart Cities:
  Improving Quality of Life Using ICT IoT (HONET-ICT)} (2017), pp.~78--82.

\bibitem[BHH16]{7827599}
\textsc{{Bayless} S., {Hoos} H.~H., {Hu} A.~J.}:
\newblock Scalable, high-quality, sat-based multi-layer escape routing.
\newblock In \emph{2016 IEEE/ACM International Conference on Computer-Aided
  Design (ICCAD)} (2016), pp.~1--8.

\bibitem[CKK19]{8648510}
\textsc{{Cho} B.~G., {Kam} D.~G., {Koo} H.~I.}:
\newblock Mixed-signal escape routing algorithm for multilayer pcbs.
\newblock \emph{IEEE Transactions on Components, Packaging and Manufacturing
  Technology 9}, 8 (2019), 1576--1586.

\bibitem[CKT{\etalchar{*}}13]{6461975}
\textsc{{Chin} C., {Kuan} C., {Tsai} T., {Chen} H., {Kajitani} Y.}:
\newblock Escaped boundary pins routing for high-speed boards.
\newblock \emph{IEEE Transactions on Computer-Aided Design of Integrated
  Circuits and Systems 32}, 3 (2013), 381--391.
\newblock \href {https://doi.org/10.1109/TCAD.2012.2221714}
  {\path{doi:10.1109/TCAD.2012.2221714}}.

\bibitem[CRN97]{10.1145/267665.267686}
\textsc{Cha Y.-J., Rim C.~S., Nakajima K.}:
\newblock A simple and effective greedy multilayer router for mcms.
\newblock In \emph{Proceedings of the 1997 International Symposium on Physical
  Design} (New York, NY, USA, 1997), ISPD '97, Association for Computing
  Machinery, p.~67–72.
\newblock URL: \url{https://doi.org/10.1145/267665.267686}, \href
  {https://doi.org/10.1145/267665.267686} {\path{doi:10.1145/267665.267686}}.

\bibitem[CWC19]{8942123}
\textsc{{Chang} Y., {Wen} H., {Chang} Y.}:
\newblock Obstacle-aware group-based length-matching routing for pre-assignment
  area-i/o flip-chip designs.
\newblock In \emph{2019 IEEE/ACM International Conference on Computer-Aided
  Design (ICCAD)} (2019), pp.~1--8.

\bibitem[DDS91]{Dai_1991}
\textsc{Dai W. W.-M., Dayan T., Staepelaere D.}:
\newblock Topological routing in surf: Generating a rubber-band sketch.
\newblock In \emph{Proceedings of the 28th ACM/IEEE Design Automation
  Conference} (New York, NY, USA, 1991), DAC ’91, Association for Computing
  Machinery, p.~39–44.
\newblock URL: \url{https://doi.org/10.1145/127601.127622}, \href
  {https://doi.org/10.1145/127601.127622} {\path{doi:10.1145/127601.127622}}.

\bibitem[Dij59]{dijkstra1959note}
\textsc{Dijkstra E.~W.}:
\newblock A note on two problems in connexion with graphs.
\newblock \emph{Numerische Mathematik 1}, 1 (1959), 269--271.

\bibitem[DKJS90]{Dai}
\textsc{{Dai} W. W.~., {Kong} R., {Jue} J., {Sato} M.}:
\newblock Rubber band routing and dynamic data representation.
\newblock In \emph{1990 IEEE International Conference on Computer-Aided Design.
  Digest of Technical Papers} (1990), pp.~52--55.

\bibitem[EKL06]{efrat2006computing}
\textsc{Efrat A., Kobourov S.~G., Lubiw A.}:
\newblock Computing homotopic shortest paths efficiently.
\newblock \emph{Computational Geometry 35}, 3 (2006), 162 -- 172.
\newblock \href {https://doi.org/https://doi.org/10.1016/j.comgeo.2006.03.003}
  {\path{doi:https://doi.org/10.1016/j.comgeo.2006.03.003}}.

\bibitem[EN11]{erickson2011shortest}
\textsc{Erickson J., Nayyeri A.}:
\newblock Shortest non-crossing walks in the plane.
\newblock In \emph{SODA '11} (2011).

\bibitem[Ful13]{fulton}
\textsc{Fulton W.}:
\newblock \emph{Algebraic topology: a first course}, vol.~153.
\newblock Springer Sci. \& Business Media, 2013.

\bibitem[HNR68]{hart1968formal}
\textsc{{Hart} P.~E., {Nilsson} N.~J., {Raphael} B.}:
\newblock A formal basis for the heuristic determination of minimum cost paths.
\newblock \emph{IEEE Transactions on Systems Science and Cybernetics 4}, 2
  (1968), 100--107.

\bibitem[HXF{\etalchar{*}}19]{8715126}
\textsc{{Huang} Y., {Xie} Z., {Fang} G., {Yu} T., {Ren} H., {Fang} S., {Chen}
  Y., {Hu} J.}:
\newblock Routability-driven macro placement with embedded cnn-based prediction
  model.
\newblock In \emph{2019 Design, Automation Test in Europe Conference Exhibition
  (DATE)} (2019), pp.~180--185.

\bibitem[JKRS94]{285679}
\textsc{{Jun Dong Cho}, {Kuo-Feng Liao}, {Raje} S., {Sarrafzadeh} M.}:
\newblock M/sup 2/r: multilayer routing algorithm for high-performance mcms.
\newblock \emph{IEEE Transactions on Circuits and Systems I: Fundamental Theory
  and Applications 41}, 4 (1994), 253--265.

\bibitem[KC93]{1600289}
\textsc{{Kei-Yong Khoo}, {Cong} J.}:
\newblock An efficient multilayer mcm router based on four-via routing.
\newblock In \emph{30th ACM/IEEE Design Automation Conference} (1993),
  pp.~590--595.

\bibitem[{Lee}61]{5219222}
\textsc{{Lee} C.~Y.}:
\newblock An algorithm for path connections and its applications.
\newblock \emph{IRE Transactions on Electronic Computers EC-10}, 3 (1961),
  346--365.

\bibitem[MCS19]{8664730}
\textsc{{Mondal} K., {Chatterjee} S., {Samanta} T.}:
\newblock An algorithm for obstacle-avoiding clock routing tree construction
  with multiple tsvs on a 3d ic.
\newblock \emph{IET Computers Digital Techniques 13}, 2 (2019), 102--109.

\bibitem[Pap96]{papadopoulou1996k}
\textsc{Papadopoulou E.}:
\newblock k-pairs non-crossing shortest paths in a simple polygon.
\newblock In \emph{Int. Symp. on Alg. and Comp.} (1996), Springer,
  pp.~305--314.

\bibitem[SMH{\etalchar{*}}21]{2021arXiv210503386S}
\textsc{{Seong} R.-K., {Min} C., {Han} S.-H., {Yang} J., {Nam} S., {Oh} K.}:
\newblock {Topology and Routing Problems: The Circular Frame}.
\newblock \emph{arXiv e-prints} (May 2021), arXiv:2105.03386.
\newblock \href {http://arxiv.org/abs/2105.03386} {\path{arXiv:2105.03386}}.

\bibitem[WHJ{\etalchar{*}}20]{Weng2020URBERUR}
\textsc{Weng J., Ho T., Ji W., Liu P., Bao M., Yao H.}:
\newblock Urber: Ultrafast rule-based escape routing method for large-scale
  sample delivery biochips.
\newblock \emph{IEEE Transactions on Computer-Aided Design of Integrated
  Circuits and Systems 39} (2020), 157--170.

\end{thebibliography}
             

\end{document}